\documentclass[aps,prl,reprint,superscriptaddress]{revtex4-1}

\usepackage{graphicx}
\usepackage{subfigure}
\usepackage{color}
\usepackage{amsmath}
\usepackage{amssymb}
\usepackage{epstopdf}
\usepackage[linktocpage,colorlinks=true,linkcolor=blue,citecolor=blue,breaklinks=true]{hyperref}
\usepackage{verbatim}
\usepackage{breakcites}
\usepackage{gensymb}
\usepackage{commath}
\usepackage[normalem]{ulem}
\usepackage[version=3]{mhchem}

\newcommand{\be}{\begin{equation}}
\newcommand{\ee}{\end{equation}}

\begin{document}

\preprint{}

\title{Seasonal evolution of the Arctic sea ice thickness distribution}

\author{Srikanth Toppaladoddi}
\affiliation{University of Leeds, Leeds LS2 9JT, U.K.}
\affiliation{University of Oxford, Oxford OX1 3PU, U.K.}

\email[]{S.Toppaladoddi@leeds.ac.uk}

\author{Woosok Moon}
\affiliation{Department of Environmental Atmospheric Sciences, Pukyong National University, 48513 Pusan, South Korea}

\author{J. S. Wettlaufer}
\affiliation{Yale University, New Haven, CT 06511, USA}
\affiliation{Nordita, Royal Institute of Technology and Stockholm University, SE-10691 Stockholm, Sweden}

\email[]{john.wettlaufer@yale.edu; john.wettlaufer@su.se}

\date{\today}

\begin{abstract}

The Thorndike et al., (\emph{J. Geophys. Res.} {\bf 80} 4501, 1975) theory of the ice thickness distribution, $g(h)$, treats the dynamic and thermodynamic aggregate properties of the ice pack in a novel and physically self-consistent manner.   Therefore, it has provided the conceptual basis of the treatment of sea-ice thickness categories in climate models.  The approach, however, is not mathematically closed due to the treatment of mechanical deformation using the redistribution function $\psi$, the authors noting ``The present theory suffers from a burdensome and arbitrary redistribution function $\psi .$''  Toppaladoddi and Wettlaufer (\emph{Phys. Rev. Lett.} {\bf 115} 148501, 2015) showed how $\psi$ can be written in terms of $g(h)$, thereby solving the mathematical closure problem and writing the theory in terms of a Fokker-Planck equation, which they solved analytically to quantitatively reproduce the observed winter $g(h)$.  Here, we extend this approach to include open water by formulating a new boundary condition for their Fokker-Planck equation, which is then coupled to the observationally consistent sea-ice growth model of Semtner (\emph{J. Phys. Oceanogr.} {\bf 6}(3), 379, 1976) to study the seasonal evolution of $g(h)$. We find that as the ice thins, $g(h)$ transitions from a single- to a double-peaked distribution, which is in agreement with observations. To understand the cause of this transition, we construct a simpler description of the system using the equivalent Langevin equation formulation and solve the resulting stochastic ordinary differential equation numerically. Finally, we solve the Fokker-Planck equation for $g(h)$ under different climatological conditions to study the evolution of the open-water fraction.
%

\end{abstract}


\maketitle

\section{Introduction}
Arctic sea ice is one of the most important components of the Earth's climate system. Its importance stems primarily from the role it plays in influencing Earth's radiation budget through its albedo and in driving the thermohaline circulation \cite{OneWatt}. Any climatological study of Arctic sea ice necessarily involves the evolution of the sea-ice volume and its interactions with the other components of the climate system. Although routine measurement of the areal extent of sea ice using satellites is now possible, a routine measurement of its thickness still remains challenging \cite{Kwok:2021}. This motivates the development of an observationally consistent mathematical theory to study the evolution of the thickness field.

The sea-ice cover consists of a complex discontinuous mosaic of floes of varying size and thickness \cite{Rothrock:1980, Rothrock:1984}, which makes any deterministic description of the system on a geophysical scale extremely difficult.  The first General Circulation Models (GCM) computed the full three-dimensional fluid dynamical and radiative-thermodynamical transport equations around an idealized globe \cite{MW75}.  However, their treatment of sea ice was simply as a thermal boundary condition on the ocean \cite{MW75}, and hence did not capture the fact that pack ice consists of a multi-scale aggregate of individual ice floes that evolve dynamically and thermodynamically.  Contemporaneously, this reality was the focus of the multi-year Arctic Ice Dynamics Joint Experiment (AIDJEX), which began in 1970 and culminated with the main field experiment from March 1975 to May 1976 \cite{Untersteiner:2007}.  In the same spirit as the GCMs the AIDJEX model treated sea ice in the spirit of weather forecasting; solving the appropriate conservation laws on a $\sim$ 100 km scale.  The momentum equation for the ice pack includes the tangential wind and ocean stresses, the Coriolis effect, and the dynamic tilt of the sea surface. Counteracting these external forces is the internal stress with which the ice pack resists deformation, parameterized by a ``constitutive law" relating the external stress to the deformation rate.  Whilst the floe-scale processes responsible for the deformation of the ice pack, and the resulting floe-size and thickness distribution, may be the foundation for the constitutive behavior, we still lack a constitutive law.  Indeed, the state of affairs during AIDJEX, as expressed by Rothrock \cite{Rothrock:1975}:
\begin{quote}
``If we knew what the constitutive equation for pack ice should be, we would not need to pay attention to the mechanisms of floe interaction. But the simple fact is that we are not at all sure about the constitutive equation...we have turned to the study of these mechanisms--rafting, ridging, shearing, and opening--to deduce what we can about the large-scale mechanical behavior of pack ice.''
\end{quote}
has not changed.  Moreover, despite having since derived a quantitative treatment of rafting and ridging \cite{VW08}, translating these into any constitutive law for the 
ice pack still poses an outstanding challenge for any continuum momentum equation that is not scale-dependent and/or highly over-parameterized for inclusion in climate models \cite{Untersteiner:2007, Coon:2007, Feltham:2008, Roberts:2019}.   

Within the AIDJEX modeling group an approach that abandons the explicit use of a momentum equation was developed by \citet{Thorndike:1975}.  They considered the corpus of mechanical, dynamical and thermodynamical effects in a region $\boldsymbol{\mathcal{R}}$, with an area $\mathcal{R}$, that give rise to the sea-ice thickness distribution, $g(h)$, defined as
\be
\int_{h_1}^{h_2} g(h) \, dh = \frac{R}{\mathcal{R}},
\label{eqn:g_defn}
\ee
where $h$ is the ice thickness and $R (\le \mathcal{R})$ is the area within $\boldsymbol{\mathcal{R}}$ that contains ice between thicknesses $h_1$ and $h_2$. 
Defined this way, $g(h)$ is the probability density function (PDF) for $h$. Their evolution equation for $g(h)$ is \cite{Thorndike:1975}
\be
\frac{\partial g}{\partial t} = - \nabla \cdot (\boldsymbol{u} \, g) - \frac{\partial}{\partial h} \left(f \, g\right) + \psi.
\label{eqn:thorndike}
\ee
Here, $\boldsymbol{u}$ is the horizontal velocity of the ice pack, $f$ is thermodynamic growth-rate of ice, and $\psi$ represents the mechanical interactions between ice floes. Note that in general the ice grows and decays and the distribution evolves under deformation and hence $g=g(h(t),t)$.  However, unless making an explicit point about this time dependence, for compactness we write $g=g(h)$, or simply $g$.

Although the concept of the ice thickness distribution has been used as an organizing principle for the categories of ice produced in momentum equation based climate models, it has not been an explicit prognostic variable, which was the original intent \cite{Thorndike:1975}.
The principal difficulty in solving Eq.~\eqref{eqn:thorndike} is associated with $\psi$, which translates the intransigence of the constitutive law problem in the momentum equation approach to the mechanical deformation in the theory for $g(h)$.  
Studies have been devoted to constructing simplified descriptions of $\psi$ \cite{Thorndike:1975, Thorndike:1992, Thorndike:2000, godlovitch2011}, but the results are either valid for only steady state or capture only the thick end of the distribution.  (Other approaches recast the work of \citet{Thorndike:1975} in different notation \cite{Horvat:2015}.) 
A more detailed discussion of these studies can be found in Toppaladoddi and Wettlaufer \cite{TW2017}.

In order to close Eq.~\eqref{eqn:thorndike} in a mathematically consistent manner, one must write $\psi$ in terms of $g(h)$.  This is done by recognizing that there is a vast separation of time and length scales between the individual mechanical interactions that change the ice thickness and the confluence of processes that change the  large-scale evolution of $g(h)$ \cite{TW2015, TW2017}.  This naturally leads to an analogy with Brownian motion, wherein there is a vast gulf between the time scale of individual collisions of solvent molecules with a pollen grain and the overall displacement of the latter.  To wit, Toppaladoddi and Wettlaufer \cite{TW2015, TW2017} interpreted $\psi$ as a collision integral:
\be
\psi(h,t) = \int_{0}^{\infty} \left[g(h^\prime,t)\, w(h,h^\prime) - g(h,t)\, w(h^\prime,h)\right] \, dh^\prime, 
\label{eqn:psi}
\ee
in which, $w(h,h^\prime)$ and $w(h^\prime,h)$ are the transition probabilities per unit time that represent deformation processes changing ice thickness from $h^\prime$ to $h$ and from $h$ to $h^\prime$, respectively.  Assuming $w(h,h^\prime) = w(h^\prime,h)$, which implies that the transition density depends only on the difference of the thicknesses of the participating ice floes, then the Kramers-Moyal-Taylor expansion of Eq. \eqref{eqn:psi} transforms Eq. \eqref{eqn:thorndike} into
\be
\frac{\partial g}{\partial t} = - \nabla \cdot ( \boldsymbol{u} \, g) - \frac{\partial}{\partial h} \left(f \, g\right) + \frac{\partial}{\partial h} (k_1 \, g) + \frac{\partial^2}{\partial h^2} (k_2 \, g),
\label{eqn:newg(h)}
\ee
where
\be
k_1 = \int_{0}^{\infty} \left|h^\prime - h\right| \, w(h,h^\prime) \, dh^\prime
\ee
and
\be
k_2 = \int_{0}^{\infty} \frac{1}{2}\left|h^\prime - h\right|^2 \, w(h,h^\prime) \, dh^\prime
\ee
are the first and second moments over the transition density \cite{TW2015, TW2017}.  The Pawula theorem \cite{Pawula} guarantees that Eq. \eqref{eqn:newg(h)} is well-posed in the sense of Hadamard, and hence obeys a maximum principle \cite{Courant}, insuring that $g(h)$ is a well defined probability density function.  Moreover, assuming that the first and second moments over the transition probabilities are constant in the region $\mathcal{R}$ is equivalent to the assumption that the physics of ice deformation is the same anywhere within it.   For example, ice ridging is governed by the same basic physical processes anywhere in the ice pack \cite{Rothrock:1975, VW08}.

Equation \ref{eqn:newg(h)} can be non-dimensionalized using $H_{eq}$, the seasonal mean ice thickness, as the vertical length scale; $L$ as the horizontal length scale; $U_0$ as the velocity scale for the horizontal ice velocity; $t_m = L/U_0$ as the time scale for advection of ice floes; $t_D = H_{eq}^2/\kappa$, where $\kappa$ is the thermal diffusivity of ice, as the diffusion time scale; and $t_R \sim 1/\dot\gamma$, where $\dot \gamma$ is the collisional strain rate, as the relaxation time scale. Because the deformation in the ice pack is driven by the wind, we have $t_m \approx t_R$. The remaining terms have the following scalings: $f_0 = H_{eq}/t_D$, $\widetilde{k_{1}} = H_{eq}/t_R$, and $\widetilde{k_{2}} = H_{eq}^2/t_R$.   Finally, observations in the central Arctic spanning 1978--2015 show that the mean divergence field is solenoidal to a part in 10$^{11}$ \cite{agarwal2017}.  Thus, we follow $\mathcal{R}$ in the Lagrangian frame, and retaining the pre-scaled notation, Eq. \eqref{eqn:newg(h)} becomes
\be
\frac{Dg}{Dt} =  \frac{\partial}{\partial h} \left[\left(k_1 - \tau f\right) g\right] + \frac{\partial^2}{\partial h^2}\left(k_2 g\right) = - \frac{\partial J}{\partial h},
\label{eqn:fpt_scaled}
\ee
where $\tau \equiv t_m/t_D \ll 1$ where
\be
J = - \left[\left(k_1 - \tau \, f \right) \, g + k_2 \, \frac{\partial g}{\partial h}\right]
\label{eqn:flux}
\ee
is the total flux of probability.  This closes the theory of \citet{Thorndike:1975}, which we have transformed into a Fokker-Planck Equation.

\begin{figure}
\centering
\includegraphics[trim = 500 0 450 0, scale=0.20]{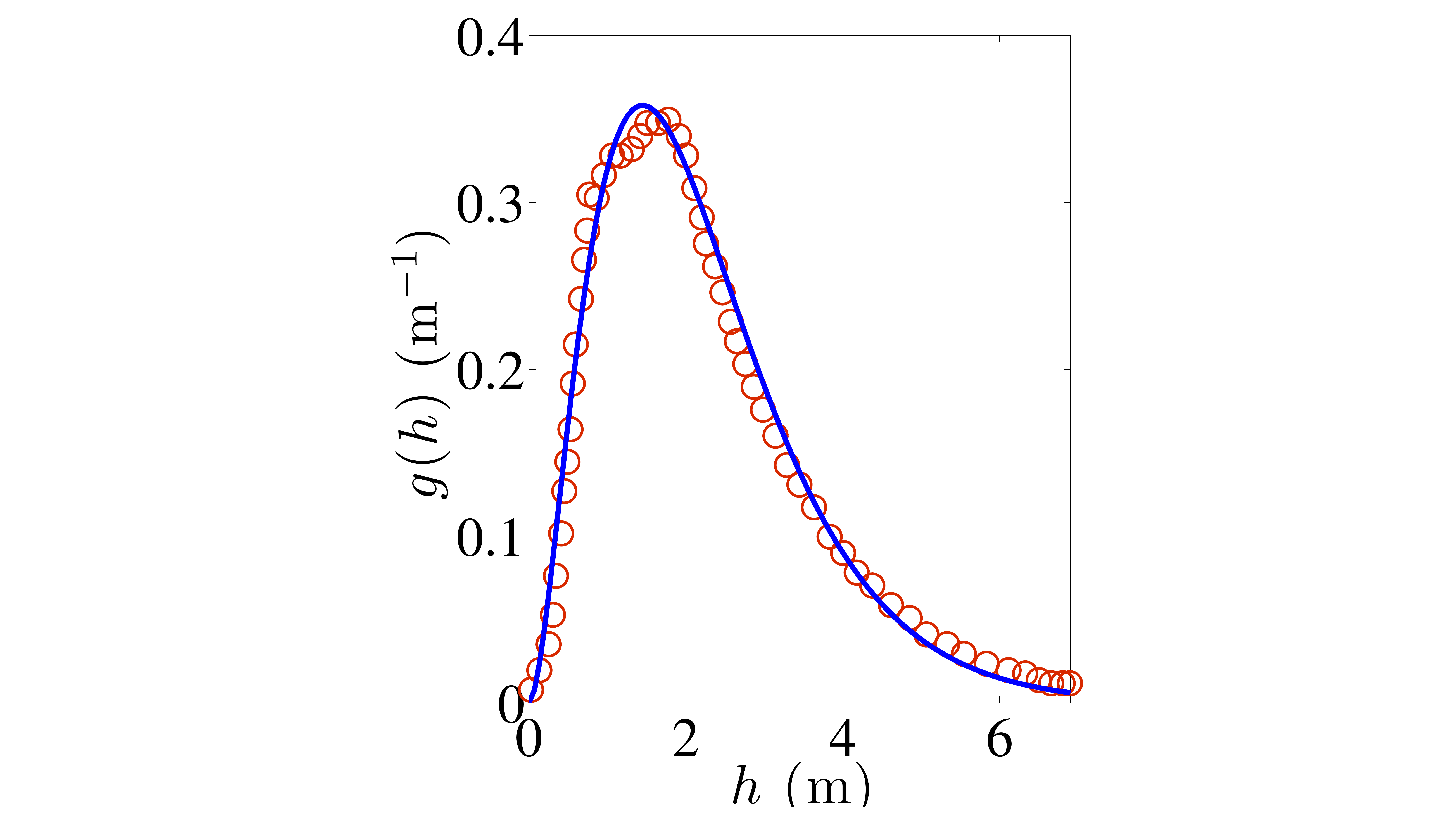}
\caption{Comparison of the steady solution to Eq. \eqref{eqn:fpt_scaled}, Eq. \eqref{eq:IM} (lines), 
with satellite measurements from ICESat \cite{kwok2009}, for February through March of 2008 (circles) from \cite{TW2015}. The two parameters from Eq. \eqref{eq:IM} are $q=1.849$ and $H=0.783 \mathrm{~m}$. The constants $k_1$ and $k_2$ are obtained using these values of $q$ and $H$.}
\vspace{-5mm}
\label{fig:prl}
\end{figure}

\textcolor{black}{During winter, open water rapidly freezes. Hence, \citet{TW2015} used the boundary conditions $g(h = 0) = g(h = \infty) = 0$ to study the winter ice pack.  They found that the winter season is governed by the following invariant measure of Eq. \eqref{eqn:fpt_scaled}, 
\be
g(h)=\mathcal{N}(q) h^{q} e^{-h / H}, 
\label{eq:IM}
\ee
with prefactor $\mathcal{N}(q)=\left[H^{1+q} \Gamma(1+q)\right]^{-1}$, wherein $\Gamma(x)$ is the Euler gamma function, 
as determined by the normalization condition $\int_{0}^{\infty} g(h) d h=1$. Hence, $\mathcal{N}(q)$ is unique and single valued for $\mathbb{R}(q)>-1$ and $\mathbb{R}(H)>0$. Here, $q=\tau c_p \Delta T/ k_2 L_i \text { and } H=k_2 / k_1$, where $L_i, c_p$ and $\Delta T$ are the latent heat of fusion of ice, the specific heat of ice at constant pressure and the temperature difference across the ice layer, respectively. 
The dimensionless thermodynamic ice growth rate is $f={c_p \Delta T}/{L_i h}\equiv {1}/{S h}$, where $S$ is the Stefan number and $k_2$ represents mechanical deformation, so that $q$ characterizes the combined effects of both processes, whereas $H$ is solely associated with mechanical deformation.  
Importantly, $q$ and $H$ are the sole parameters associated with the bivariate satellite observations for the winter months \cite{kwok2009, kwok2015} as reproduced here in Fig.~\ref{fig:prl}.}  Thus, for $h \ll 1, g(h)$ is controlled by both thermodynamics and mechanics, whereas for $h \gg 1, g(h)$ is controlled solely by mechanical interactions, showing that the thick end of the distribution can only be achieved by ice deformation. The dimensionless constants $k_1$ and $k_2$ are obtained using:
\[q = \frac{\epsilon}{k_2} \hspace{0.2cm} \text{and} \hspace{0.2cm} H = \frac{k_2}{k_1},\]
where $\epsilon = \tau/S$. We estimate $\tau \approx 0.46$ and $S \approx 10$, giving $\epsilon \approx 0.046$, and using $H_{eq} = 1.5$ m, we get $k_1 = 0.048$ and $k_2  = 0.025$. These are the values used in this study.
%

We note that the condition $g(h=0) = 0$ implies that there is no open water in the study region $\mathcal{R}$. As shown in Fig.~\ref{fig:prl}, this is a reasonable approximation in winter, 
when the open water rapidly freezes and on average only a small fraction of the distribution neglected.  However, this is not the case for the full seasonal cycle.  Indeed, a particular challenge in formulating a boundary condition for $g(h)$ at $h=0$ is that open water forms through both thermodynamic and mechanical processes.  Thus, $g(h=0)$ must be obtained as a part of the solution to Eq. \eqref{eqn:fpt_scaled}.  Therefore, here we formulate a complete seasonal boundary condition for $g(h)$ at $h=0$ and study the evolution of $g(h)$ with the aide of the one-dimensional sea-ice growth model of \citet{semtner1976} in Eq. \eqref{eqn:fpt_scaled}.

%

\section{The Open Water fraction }

Let $A$ be the fraction of open water present in the region $\mathcal{R}$. The normalization condition for $g(h)$ is then 
\be
A + \int_{0^+}^{\infty} g(h,t) \, dh = 1.
\label{eqn:normalization}
\ee
Differentiating this with respect to $t$ gives
\be
\frac{dA}{dt} = - \int_{0^+}^{\infty} \frac{D g(h,t)}{D t} \, dh, 
\label{eqn:normalization2}
\ee
and using Eq. \eqref{eqn:fpt_scaled} in Eq. \eqref{eqn:normalization2} yields 
\be
\frac{dA}{dt} =  \left[J|_{h = \infty} - J|_{h=0^+}\right] = - J|_{h=0^+}.
\label{eqn:openwater}
\ee
Therefore, as is observed, $A$ increases (decreases) as the Arctic enters spring and summer (fall and winter) during which $J|_{h=0^+} \le 0$ ($J|_{h=0^+} \ge 0$).



In order to relate the open water fraction, $A(t)$, to the thickness distribution at the origin, $g(h=0,t)$, we let:
\be
A(t) \equiv g(0,t) \, H_c. 
\label{eqn:cutoff}
\ee
In dimensional terms $\tilde{H}_c \equiv \zeta \Lambda$, where $\zeta$ is the fraction of the \textcolor{black}{spectrally} and angularly averaged Beer's extinction length, $\Lambda$, below which thin ice and open water become indistinguishable.  Taking $\zeta$ to be 15\% of the extinction length appropriate for $H_{eq}$, which is $\Lambda$ = 67 cm \cite{MU71}, gives $\tilde{H}_c$ = 10 cm.  
Using this in Eq. \eqref{eqn:openwater} gives 
\be
\frac{dg(0,t)}{dt} = - \frac{1}{H_c} \, J|_{h=0^+},
\label{eqn:new_bc}
\ee
which is the required evolution equation for $g(0,t)$. Equation \eqref{eqn:fpt_scaled}, along with the boundary conditions \eqref{eqn:new_bc} and $g(\infty) = 0$, can now be used to solve for $g(h,t)$. Once $g(h,t)$ is known, $A(t)$ can be calculated from Eq. \eqref{eqn:normalization}.

We follow \citet{TW2015} and impose $g(0,t) = 0$ during winter, and equation \eqref{eqn:new_bc} for the remaining part of the year. The transition between these boundary conditions is determined by the sign of the thermal growth rate of open water, $f(0,t)$, which is positive in winter.

\section{Thermal growth of sea ice}

To calculate the thermal growth rate of sea ice, $f(h,t)$, we use the observationally consistent one-dimensional model by \citet{semtner1976}. The thickness of the snow layer is assumed to be uniform across all the thicknesses and is prescribed following \citet{MU71}: 30 cm from August 20 to October 30, 5 cm from November 1 to April 30, and 5 cm during the month of May. Snow is taken to accumulate only when the mean surface temperature of the ice layer or the snow layer is below the freezing point, and the increase in the snow thickness is taken to be linear \cite{MU71}. The values of snow albedo for the different months are taken from \citet{MU71}, and the thickness-dependent albedo of sea ice is obtained using the expression from \citet{EW09}:
\be
\alpha(h) = \left(\frac{\alpha_w + \alpha_i}{2}\right) + \left(\frac{\alpha_w-\alpha_i}{2}\right) \, \tanh\left(-\frac{h}{\Lambda}\right),
\label{eqn:albedo}
\ee
\textcolor{black}{where $\alpha_w$ and $\alpha_i$ are the albedos of open water and the thickest ice, respectively, and as discussed above $\Lambda$ is the Beer's extinction length for ice. Furthermore, in the absence of snow, the fraction of the net shortwave radiation that penetrates ice is taken to be 17\% \cite{MU71}. In addition to the shortwave radiation, incoming longwave radiation, and specific and latent heat fluxes from the atmosphere, we also include a perturbation to the incoming longwave radiation, $\Delta F_0$, which represents the effects of addtional greenhouse gas forcing. }

Semtner's numerical formulation does not permit the inclusion of an internal heat source that represents the penetration of shortwave radiation \cite{semtner1976}. Rather, this energy is stored in a ``reservoir" and released during the fall freeze-up. The latent heat of fusion at the top surface is adjusted when the energy in the reservoir exceeds 30\% of what is required to melt the entire ice layer \cite{semtner1976}. When the thickness distribution is considered, this formulation makes the latent heat of fusion a function of the ice thickness, which is unphysical. Hence, we ignore this storage effect and note that it only impacts the time at which temperature of the upper ice surface drops below the freezing point in fall. We note here that neglecting brine pockets and this heat source in the ice has implications for energy conservation, especially for long-time simulations \cite{bitz1999}. The three-layer Semtner model employed here is adequate to represent the effects of the snow cover and the specific heat of sea ice for our purposes.  However, we emphasize that it is simple to use other thermodynamic models in this framework as we have shown previously \cite{TW2017}.

\section{Results}

\textcolor{black}{We solve Eq. \eqref{eqn:fpt_scaled} numerically using a flux-conserving, fully implicit, finite-difference scheme, subject to boundary conditions \eqref{eqn:new_bc} and $g(\infty) = 0$ (or $g(0) = 0$ and $g(\infty) = 0$). The growth and decay of open water is accompanied by the evolution of a boundary layer at $h=0$, the resolution of which requires a finer numerical grid than in our previous work \cite{TW2017}. The radiative fluxes used to compute the growth rate are taken from observations \cite{MU71}.  Finally, unless otherwise stated, $\Delta F_0$ is set to  $0$ Wm$^{-2}$ and $F_B$, the oceanic heat flux, is set to $2$ Wm$^{-2}$.}
%

The results presented in the following sections were obtained after the system reached a statistically steady state. 

\subsection{Seasonality of the ice thickness distribution}
We first focus on the evolution of $g(h)$ during a typical year. 
\begin{figure}
\centering
\includegraphics[trim = 100 0 100 0, clip, width = 1\linewidth]{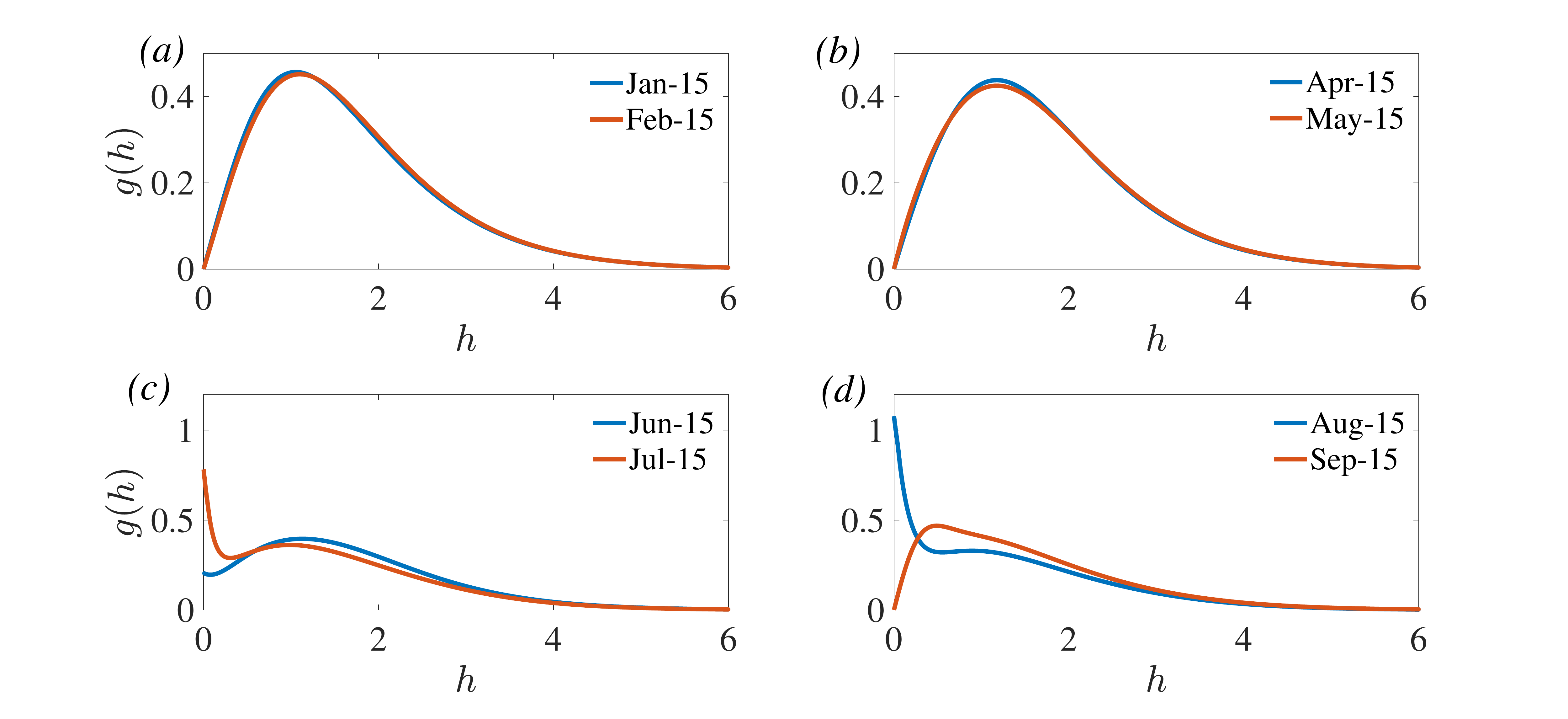}
\caption{Thickness distributions during different months for $F_B = 2$ Wm$^{-2}$ and $\Delta F_0 = 0$. The second peak in $g(h)$ emerges around June 15, which is seen in figure (c).}
\label{fig:gseasons}
\end{figure}
\textcolor{black}{Figures \ref{fig:gseasons}(a) -- \ref{fig:gseasons}(d) show that during winter (summer), $g(h)$ expands (contracts), in agreement with observations \cite{kwok2015}. The key features are as follows: (i) The distribution for June 15 in figure \ref{fig:gseasons}(c) is double-peaked; (ii) The value of the vertical intercept, $g(0)$, increases from June to August, as seen in figure \ref{fig:gseasons}(d); and (iii) as fall begins, this double-peaked distribution evolves to a single-peaked distribution in September, as shown in figure \ref{fig:gseasons}(d). These distributions should be contrasted with the those obtained by Toppaladoddi and Wettlaufer \cite{TW2017} using $g(0) = 0$ for the entire season. }

\begin{figure}
\centering
\includegraphics[trim = 0 0 0 0, clip, width = 1\linewidth]{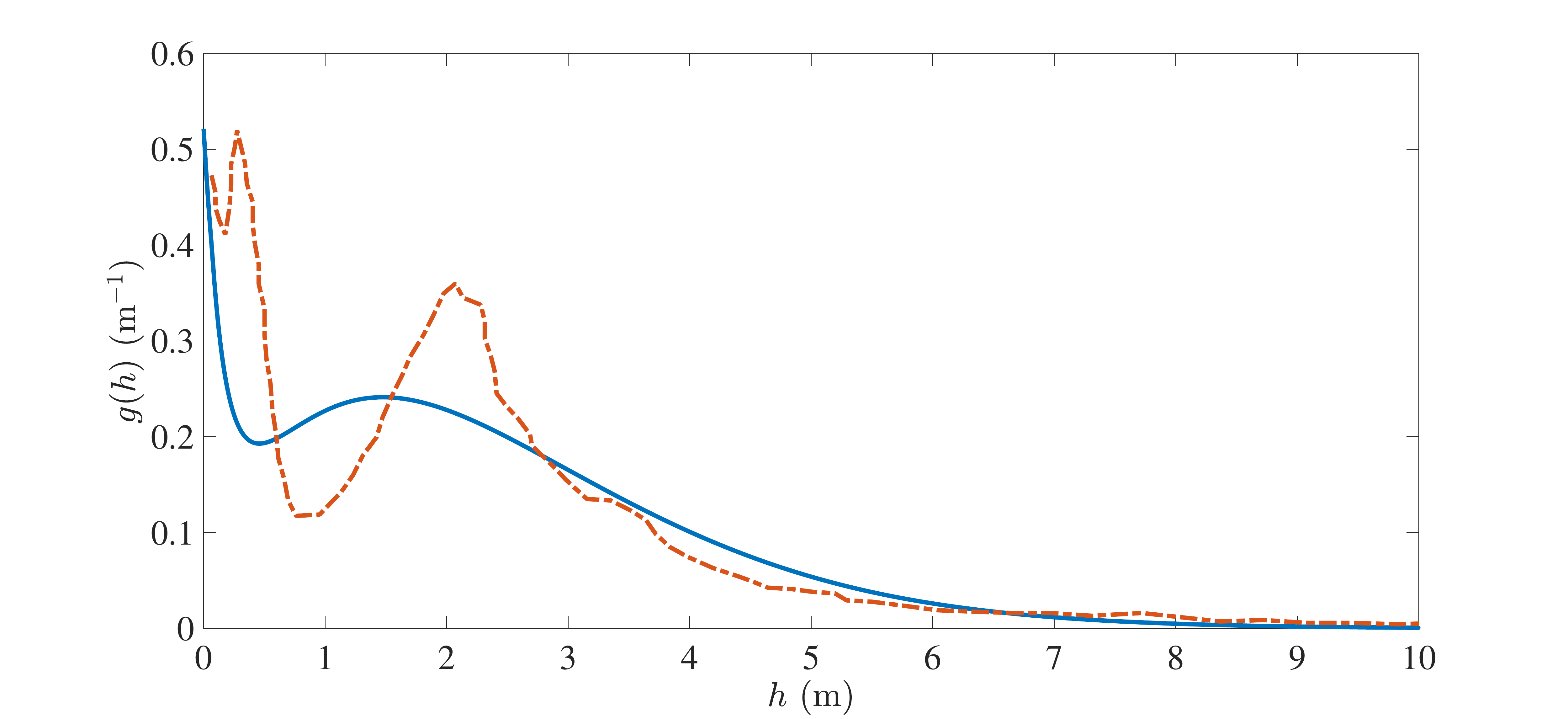}
\caption{Qualitative comparison of doubled-peaked $g(h)$ with observations. The solid line represents $g(h)$ for July 15 obtained from theory, and the dashed line represents the $g(h)$ from SCICEX measurements in September 1993 averaged over the entire cruise track in the Arctic \cite{yu2004}. The $g(h)$ from the theory has been made dimensional by scaling with $H_{eq}$, and the original distribution for draft from \citet{yu2004} has been converted to that for thickness and normalized.}
\label{fig:rothrock_compare}
\end{figure}

Of particular interest is the double-peaked $g(h)$ obtained here, which is a much sought after observational feature of the seasonal ice-pack.  For example, 
similar profiles for the probability density function of draft (the thickness of the submerged portion of sea ice) have been obtained using upward looking sonar measurements from submarines \cite{yu2004}. These profiles have been converted to the sea-ice thickness and have been normalized. Figure \ref{fig:rothrock_compare} shows a qualitative comparison between $g(h)$ for July 15 obtained from theory and the $g(h)$ from submarine measurements from SCICEX cruises, which were made during September 1993 with the data averaged over the entire cruise track \cite{yu2004}. The agreement between theory and observations is evident, with the timing of the onset of the second peak being controlled by the heat capacity of ice and the snow layer. The observations in figure \ref{fig:rothrock_compare}(b) can also be compared with the distribution on August 15 in figure \ref{fig:gseasons}(d), which shows a distinct second peak.
Note that, relative to winter, the summer satellite sea-ice thickness record from CryoSat-2 has many sources of variability \cite{Landy2022}.

\subsection{Evolution of the open-water fraction}

\begin{figure}
\centering
\includegraphics[trim = 0 0 0 0, clip, width = 1\linewidth]{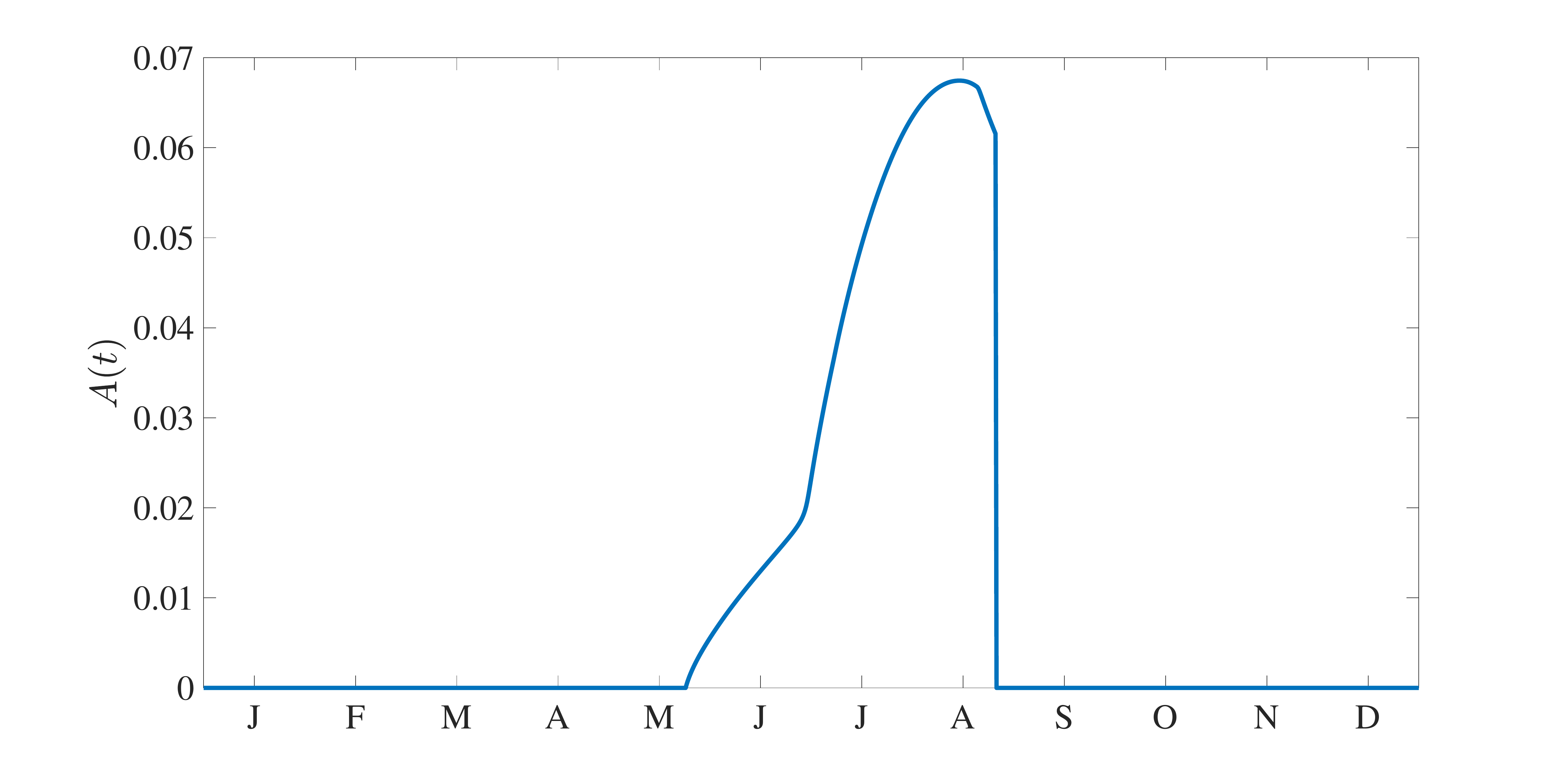}
\caption{Evolution of $A(t)$ during a typical year for $F_B = 2$ Wm$^{-2}$ and $\Delta F_0 = 0$.}
\label{fig:open_water}
\end{figure}

\textcolor{black}{The seasonal evolution of the open water fraction, $A(t)$, shown in figure \ref{fig:open_water}, constitutes a key aspect of the evolution of $g(h)$ shown in figure \ref{fig:gseasons}.}
\textcolor{black}{From the fall freeze-up through May the open water fraction is nearly zero, after which it starts to increase appreciably. The maximum value of $A$ ($\approx 6.7 \%$) is attained in mid-August. This qualitative behaviour of $A(t)$ is clearly in accord with both intuition and, most importantly, large-scale observations in the Arctic.}

\subsection{Mean thickness and albedo}

\begin{figure}
\centering
\includegraphics[trim = 50 0 50 0, clip, width = 1\linewidth]{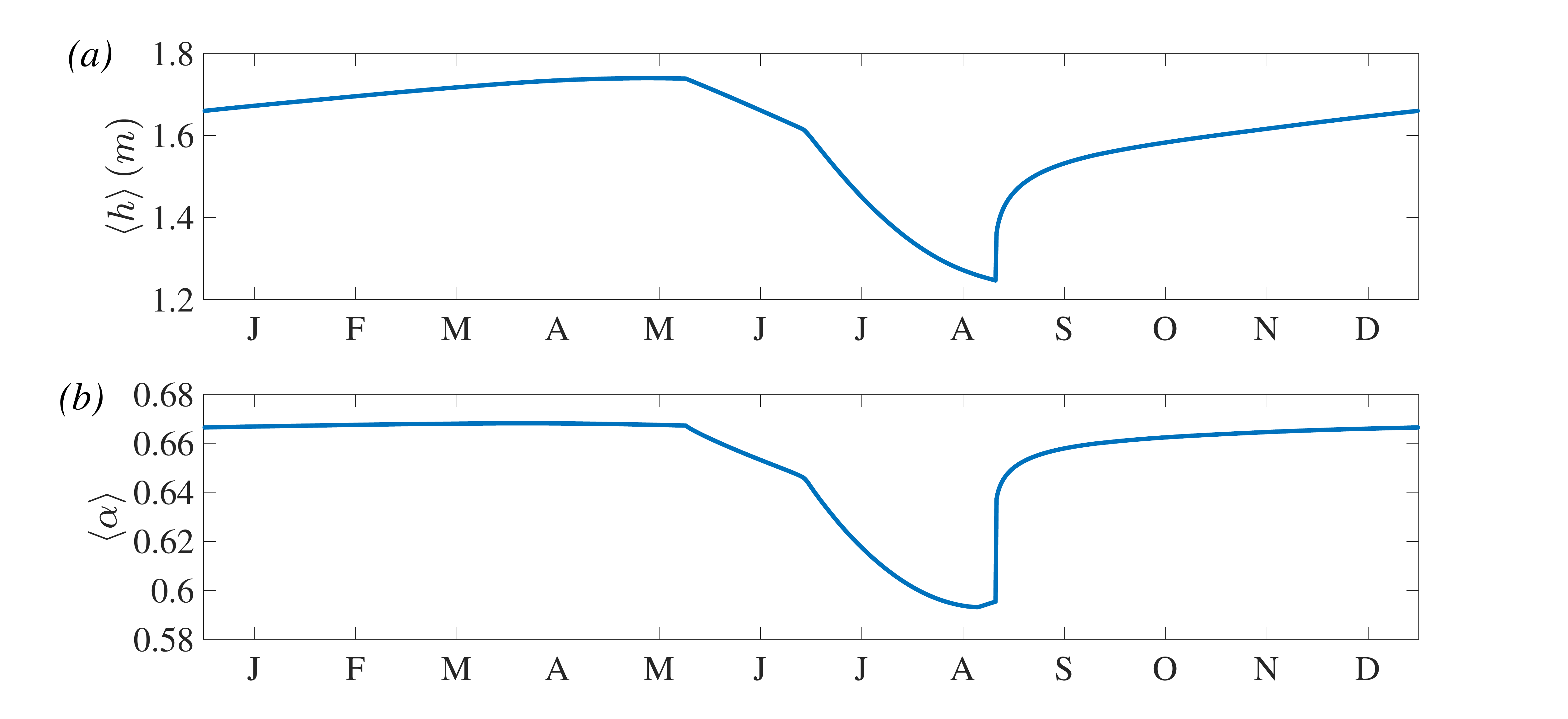}
\caption{Evolution of (a) mean thickness, $\left<h\right>$; and (b) mean albedo, $\left<\alpha\right>$ for $F_B = 2$ Wm$^{-2}$ and $\Delta F_0 = 0$.}
\label{fig:mean}
\end{figure}

The mean of any thickness-dependent quantity, $\Phi(h)$, is given by
\be
\left<\Phi(t)\right> = \int_0^{\infty} \Phi(h) \, g(h,t) \, dh.
\ee
Using this relation we calculate $\left<h\right>$ and $\left<\alpha\right>$, whose seasonal cycles are shown in figure \ref{fig:mean} for $F_B = 2$ Wm$^{-2}$ and $\Delta F_0 = 0$.
Clearly $\left<h\right>$ reaches a maximum ($\approx 1.74$ m) in the last week of May and a minimum ($\approx 1.25$ m) in the last week of August. These values are lower than for the case when $g(0) = 0$ is imposed as a boundary condition throughout the year \cite{TW2017}. A similar change can also be seen in $\left<\alpha\right>$ which here varies between $0.668$ in early April to $0.593$ in mid-August. 
\textcolor{black}{Note that $\left<\alpha\right>$ leads $\left<h\right>$, underlying the ice-albedo feedback.}

\subsection{Effects of ocean heat flux and greenhouse-gas forcing on the open-water fraction and the mean thickness}

\begin{figure}
\centering
\includegraphics[trim = 0 0 0 0, clip, width = 1\linewidth]{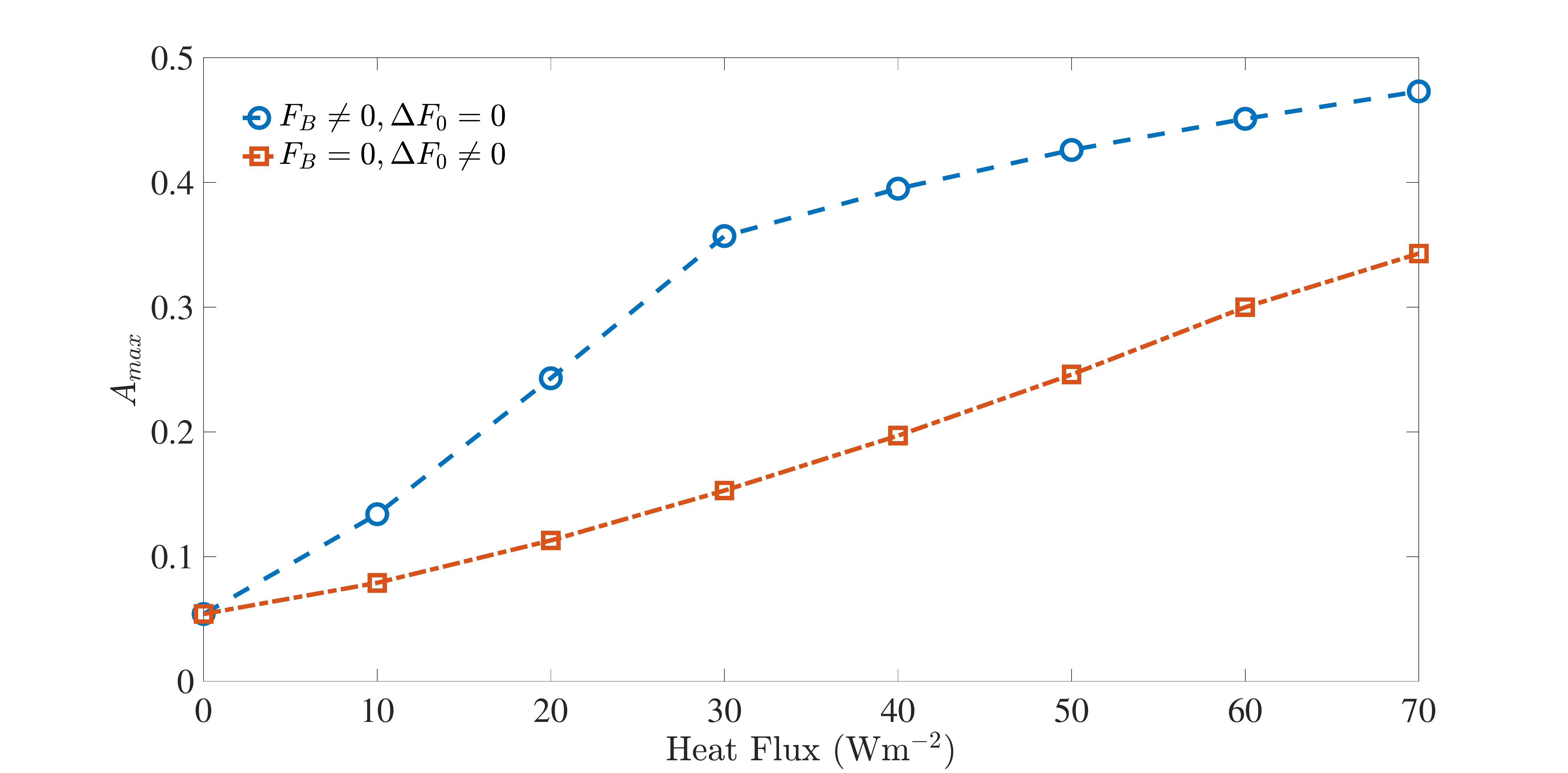}
\caption{Changes in the maximum value of the open-water fraction, $A_{max}$, with $F_B$ and $\Delta F_0$.}
\label{fig:GHG}
\end{figure}

Figure \ref{fig:GHG} shows the changes in the maximum value of open-water fraction, $A_{max}$, for the different values of $F_B$ and $\Delta F_0$. 
\textcolor{black}{Two main features are apparent: First, an increase in the value of $F_B$ and/or $\Delta F_0$ leads to an increase in the value of $A_{max}$. Second, the increase in $A_{max}$ due to $F_B$ is more rapid than that for $\Delta F_0$. Thus the ice cover is more sensitive to the ice-ocean heat flux than it is to the greenhouse-gas forcing at the upper surface. This is due to the fact that in a thermodynamic model when the Stefan number is large the temperature gradient in the ice is quasi-steady and linear \cite[e.g.,][]{EW09}.  Thus both $\Delta F_0$ and $F_B$ have the same impact on the ice thickness, whereas in a model with curvature in the temperature field these forcings are local. This sensitivity of the ice cover to the ocean heat flux is also in qualitative agreement with the results from the thermodynamic-only full heat conduction model of \citet{MU71}.}

\begin{figure}
\centering
\includegraphics[trim = 0 0 0 0, clip, width = 1\linewidth]{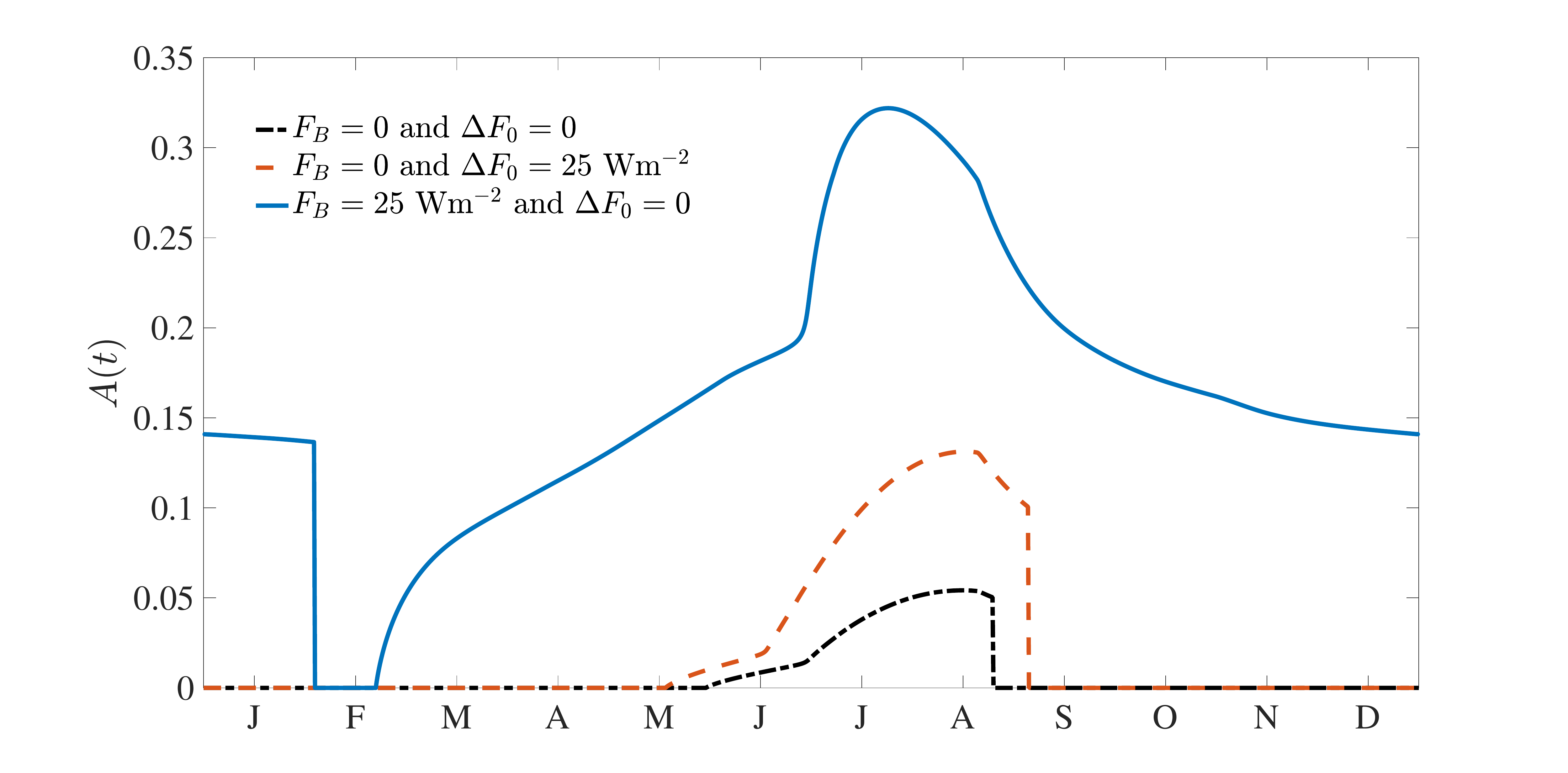}
\caption{Changes in $A(t)$ for different values of $F_B$ and $\Delta F_0$.}
\label{fig:GHG_FB}
\end{figure}

\textcolor{black}{Not only do $F_B$ and $\Delta F_0$ impact the maximum values of $A(t)$, but also its minimum values and the time at which open water starts forming, as shown in figure \ref{fig:GHG_FB}.}
When $\Delta F_0 = 25$ Wm$^{-2}$ and $F_B = 0$, which corresponds to a six-fold increase in the CO$_2$ concentration in the atmosphere, open water begins to form near the middle of May and persists until early September. This is due to the fact that the increase in $\Delta F_0$ leads to an earlier disappearance of the snow layer, and hence earlier melting of the ice layer. These changes can be clearly contrasted with the case when both $\Delta F_0$ and $F_B$ are set to zero. Importantly, we note that the choice of $\Delta F_0 = 25$ Wm$^{-2}$ and $F_B = 0$ has been made to highlight the effects of these fluxes on the ice cover. Clearly, when there is a substantial increase in the atmospheric CO$_2$ concentration the oceanic heat flux will increase.

\begin{figure}
\centering
\includegraphics[trim = 0 0 0 0, clip, width = 1\linewidth]{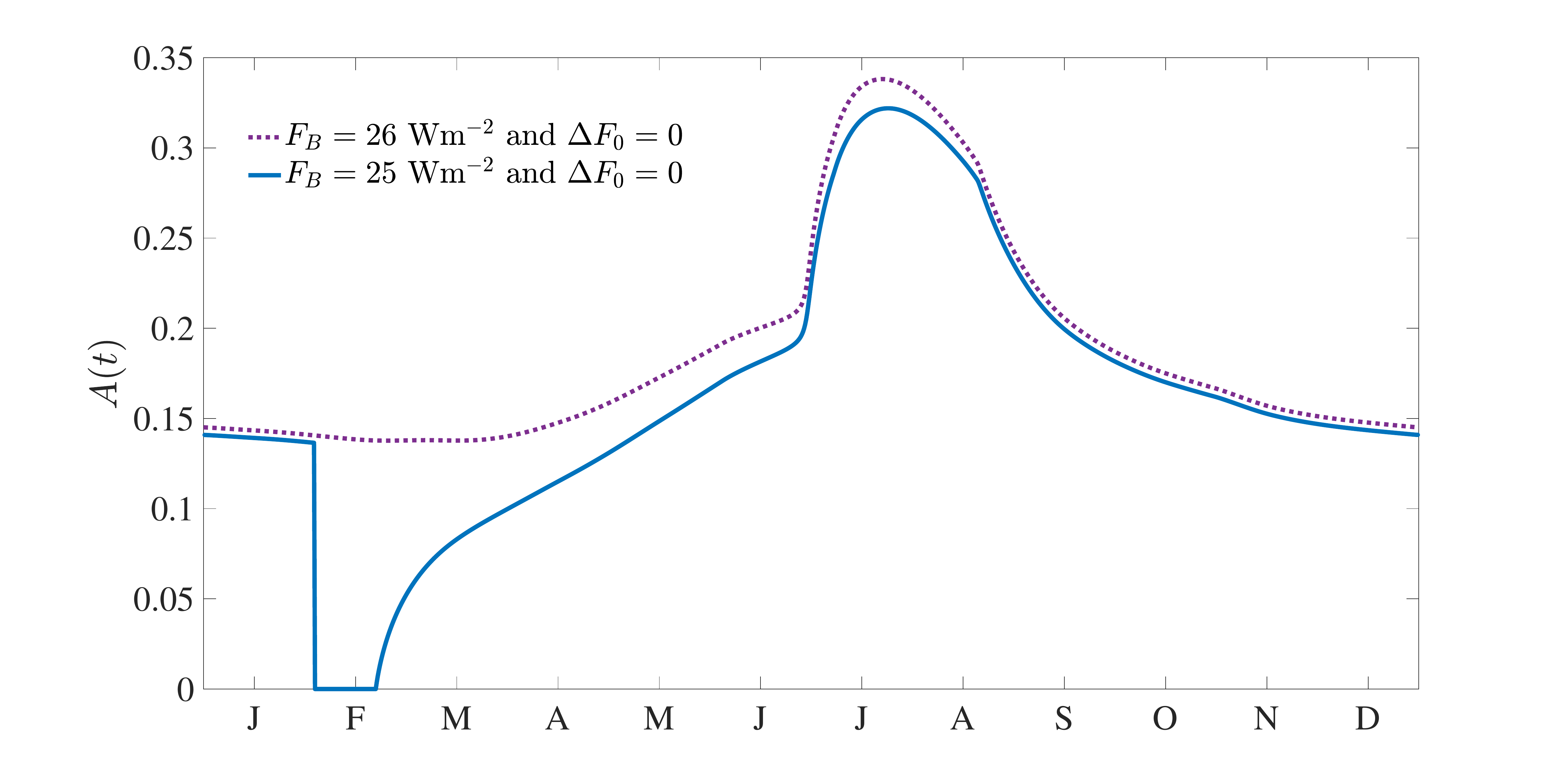}
\caption{Changes in $A(t)$ for $F_B = 25$ and $26$ Wm$^{-2}$.}
\label{fig:FB_sensitivity}
\end{figure}

\begin{figure}
\centering
\includegraphics[trim = 0 0 0 0, clip, width = 1\linewidth]{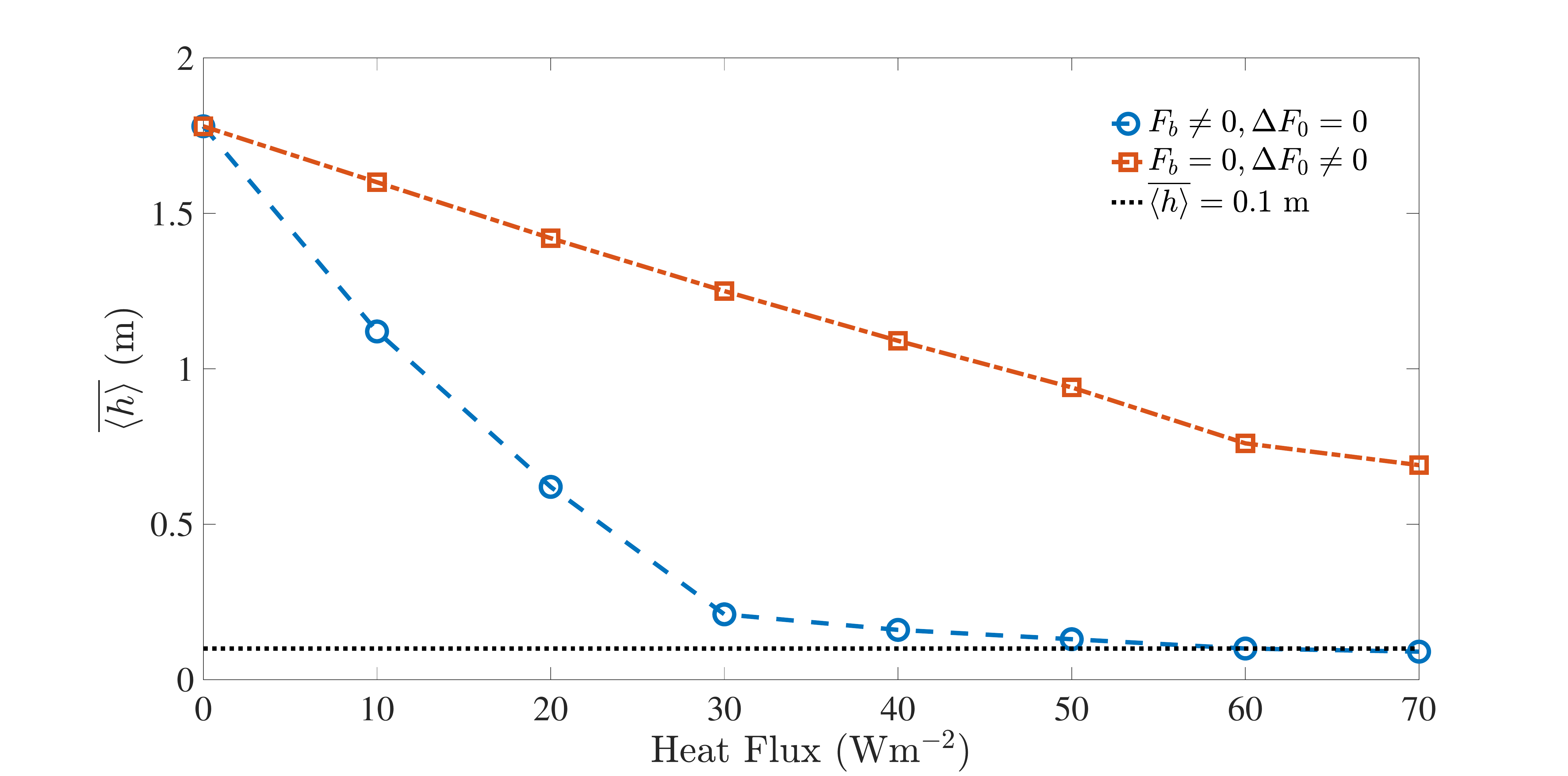}
\caption{Changes in the time-averaged mean thickness, $\overline{\left<h\right>}$, with $F_B$ and $\Delta F_0$.}
\label{fig:mean_thickness}
\end{figure}

The changes in $A(t)$ due to $F_B$ are more striking. \textcolor{black}{When $F_B = 25$ Wm$^{-2}$ (and $\Delta F_0 = 0$), $A_{max}$ is about $32 \%$, and open water is present throughout nearly the entire year. Moreover, the change in the energy balance is such that thin ice only grows from open water during only three weeks in February. To further highlight the sensitivity of the ice cover to $F_B$, we show in figure \ref{fig:FB_sensitivity} the evolution of $A(t)$ for $F_B = 25$ and $26$ Wm$^{-2}$. An increase in $F_B$ of just $1$ Wm$^{-2}$ results in open water throughout the year. These effects are due to the thickest ice being ostensibly isothermal at the base and hence any non-zero value of $F_B$ drives ablation \cite{MU71}. This leads to a greater thinning of the ice cover (see figure \ref{fig:mean_thickness}).}

Similar effects of $F_B$ and $\Delta F_0$ are also seen in the changes to the time-averaged mean thickness, $\overline{\left<h\right>}$, which are shown in figure \ref{fig:mean_thickness}.
Here, the time average is taken over a typical year in the statistically steady state. The decay in $\overline{\left<h\right>}$ with $F_B$ is exponential, whereas it is algebraic with $\Delta F_0$. When $F_B = 60$ Wm$^{-2}$, $\overline{\left<h\right>} = 0.1$ m = $H_c$, which implies that it is difficult to distinguish between thin ice and open water. The value of the basal heat flux for which nearly ice-free conditions are observed with $g(h)$ is an order of magnitude larger than seen in the calculations of \citet{MU71}, \textcolor{black}{but is consistent with more recent measurements made in the Arctic \citep{JSW91, mcphee1992, stanton2012}}. \textcolor{black}{This persistence of the ice cover has two principal causes both captured by our theory: The thickest ice survives the melt season and thinner ice rafts and ridges to become thicker and hence has a higher albedo through Eq. \eqref{eqn:albedo}.}



\subsection{Transition from a single- to a double-peaked distribution}

\textcolor{black}{To examine the transition of $g(h)$ from being single- to double-peaked, we use the equivalent Langevin formulation corresponding to Eq.~\eqref{eqn:fpt_scaled}, which is }
\be
\frac{dh}{dt} = \left(\tau \, f  - k_1 \right) + \sqrt{2 \, k_2} \, \xi(t),
\label{eqn:langevin}
\ee
where $\left(\tau \, f - k_1 \right)$ and $\sqrt{2 \, k_2} \, \xi(t)$ are the drift and diffusion terms, respectively, and $\xi(t)$ is Gaussian white noise \cite{TW2015}. The growth rate here is taken to be
\be
f = \frac{1}{\rho_i \, L_i \, f_0} \, \left(k_i \, \frac{\Delta T}{h} - F_B \right),
\label{eqn:spring_growth}
\ee
\textcolor{black}{where $k_i$ is the thermal conductivity of ice, $\Delta T$ is the temperature difference across the ice layer, and, as we have done throughout, we assume $F_B$ to be a constant.} 


\textcolor{black}{An ensemble of $N_{en} \approx 10^5$ thicknesses constitute the initial conditions for Eq.~\eqref{eqn:langevin}, such that their distribution corresponds to the winter solution of Eq.~\eqref{eq:IM}. For each realization Eq.~\eqref{eqn:langevin} is then integrated for $\mathcal{T} = 1.25$ in non-dimensional units. The behavior at the origin is treated by requiring that if $h < 0$ at the end of integration in any realization, then $h$ is set to $0$. }
\begin{figure}
\centering
\includegraphics[trim = 50 0 50 0, clip, width = 1\linewidth]{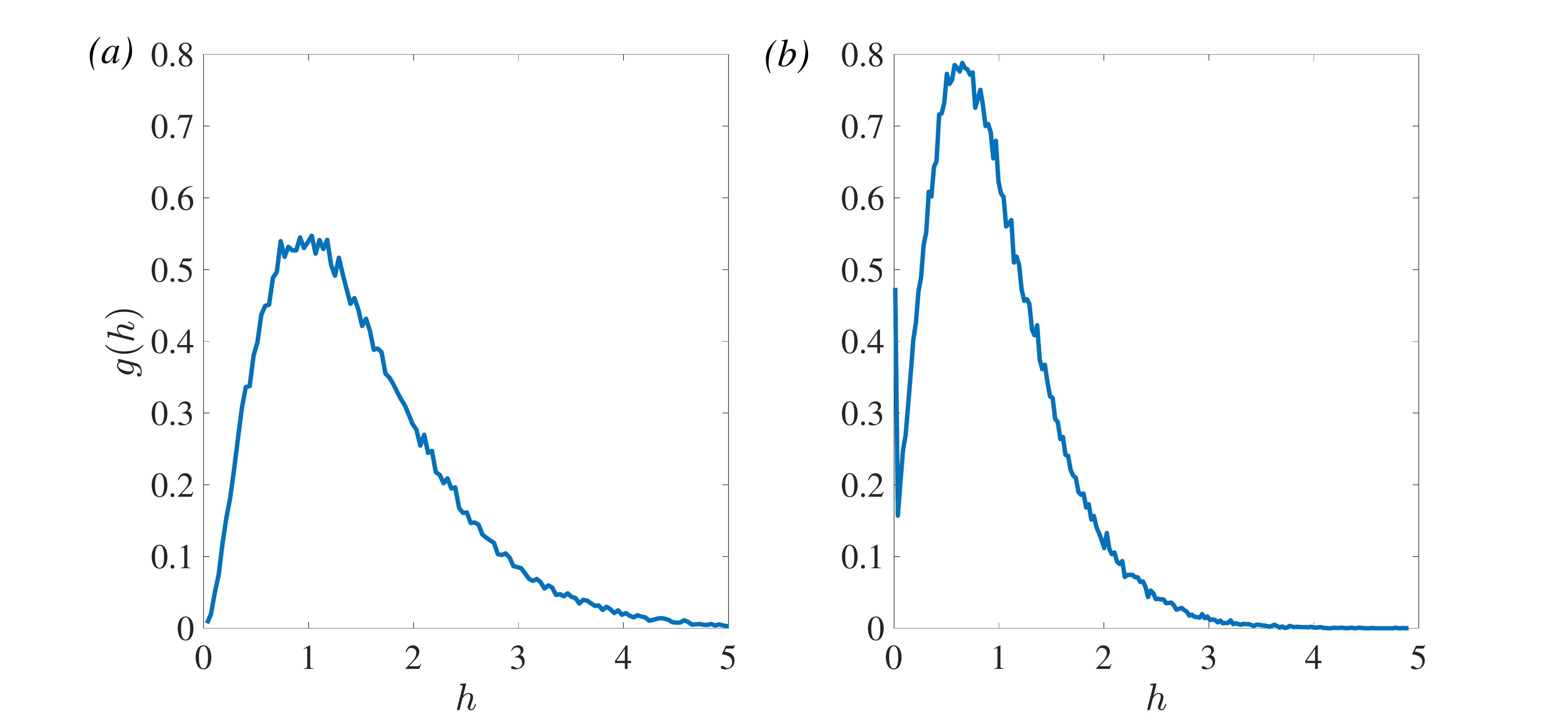}
\caption{Distributions obtained from solving Eq.~\eqref{eqn:langevin} for $F_B = 5$ Wm$^{-2}$ and $\Delta T = 5$ \degree C. Here, $N_{en} \approx 10^5$, $\mathcal{T} = 1.25$, and $\Delta t = 5 \times 10^{-5}$. Figure (a) shows the the initial $g(h)$ and figure (b) shows the $g(h)$ at the end of integration.}
\label{fig:langevin}
\end{figure}

\textcolor{black}{The transition in $g(h)$ to a double-peaked distribution is shown in figure \ref{fig:langevin}.  Clearly, as the energy balance on the right hand side of Eq.~\eqref{eqn:spring_growth} changes, so too does the sign of the drift term in Eq.~\eqref{eqn:langevin}.  Therefore, the principal factor responsible for the emergence of the double-peaked distribution is the change in energy balance.}

\section{Conclusions}
We have closed the original theory of the sea ice thickness distribution of \citet{Thorndike:1975} by recasting the mechanical redistribution function in terms of the probability density of ice thickness itself; $g(h)$.
\textcolor{black}{In consequence of this closure the original theory becomes a Fokker-Planck equation.  
We then generalized the theory to include the seasonal variation in the open water fraction. This is achieved by formulating a new boundary condition  for $g(h)$ at $h = 0$. The numerical solutions show a transition from a single-peaked to a doubled-peaked $g(h)$ in summer, in general agreement with submarine measurements from SCICEX cruises made in 1993 \cite{yu2004}. }

\textcolor{black}{Our formulation makes the explicit calculation of the open-water fraction, $A(t)$, possible. 
In the absence of excess greenhouse-gas forcing, $\Delta F_0 = 0$, the solutions reveal that $A(t) \approx 0$ during the winter months, increases near the end of May and persists until the end of August. However, an increase in $\Delta F_0$ to $25$ Wm$^{-2}$ shifts the onset of open-water formation to the middle of May and it persists until early September.}

\textcolor{black}{Particularly dramatic are the effects of the basal heat flux, $F_B$, on the ice cover.  The time-averaged mean thickness decreases exponentially with increasing $F_B$, and when $F_B = 26$ Wm$^{-2}$, open water persists for the entire year. However, the value of $F_B$ for which nearly ice-free conditions are observed with $g(h)$ is an order of magnitude larger than seen in the thermodynamic-only calculations \citep{MU71, semtner1976, EW09}. This is because in our theory for $g(h)$ the thickest ice survives the melt season and thinner ice rafts and ridges to become thicker and hence has a higher albedo.  The basal heat flux is thus far more impactful than is greenhouse-gas forcing, $\Delta F_0$.}

\textcolor{black}{These results also highlight the different roles played by thermodynamics and mechanics in the evolution of the ice cover as the Arctic ocean warms \citep{MLT2015, MLT2018}. An increase in the ocean heat flux tends to make the ice cover thinner, but the mechanical redistribution of ice prolongs its survival.}

The necessity of including the thickness distribution into global and regional climate models to accurately capture the atmosphere-ice-ocean interactions has long been recognized in the sea-ice modelling community \cite{hibler1979, bitz2001, smith2022}. The thickness distribution is generally resolved using five thickness categories (excluding open water) and the treatment of the mechanical redistribution is based on the original formulation of \citet{Thorndike:1975}, which -- according to \citet{Thorndike:1975} -- is ``arbitrary". The principal source of this arbitrariness comes from the need to specify the range of ice thicknesses on the thin end of the thickness distribution that ridge to form thicker ice. However, in principle, ice of all thicknesses can participate in ridging \citep{VW08}.

The inclusion of the thickness distribution is crucial for improving the energy balance in climate models.  However, the arbitrariness associated with the mechanical redistribution of sea ice is a possible reason why some climate models include a thickness distribution that does not evolve in time \cite{castro2014}. Indeed, despite advances in sea-ice modelling, the CMIP models do not realistically capture the observed spatial distribution of sea ice thickness or ice extent \cite{stroeve2014, agarwal2018, wei2020}. Hence, we believe that progress can be made by addressing the issue of the arbitrariness associated with the mechanical redistribution of sea ice.

\textcolor{black}{Our theory does not suffer from this arbitrariness and allows for mechanical interactions in ice of any thickness. We show that the theory provides a physically robust and observationally consistent framework to study the seasonal evolution of the thickness distribution. 
Moreover, our prediction of the emergence of a second peak in the distribution, which is a key aspect of observations, highlights the necessity of  the higher resolution required to capture the behavior near $h=0$ in summer (figure \ref{fig:rothrock_compare}(a)). Finally, it is hoped that our work will lead to a more realistic representation of the thickness distribution in climate models, and hence to more accurate predictions of the fate of the Arctic ice cover.}

\section*{Acknowledgements}
The authors acknowledge the support of the Universities of Leeds and Oxford, NORDITA, and Yale University.
S.T. acknowledges a Research Fellowship from All Souls College, Oxford and helpful discussions with A. J. Wells.  W.M. and J.S.W. acknowledge Swedish Research Council grant no. 638-2013-9243 for support. Nordita is partially supported by Nordforsk.

\bibliography{g(h)}

\begin{thebibliography}{41}%
\makeatletter
\providecommand \@ifxundefined [1]{%
 \@ifx{#1\undefined}
}%
\providecommand \@ifnum [1]{%
 \ifnum #1\expandafter \@firstoftwo
 \else \expandafter \@secondoftwo
 \fi
}%
\providecommand \@ifx [1]{%
 \ifx #1\expandafter \@firstoftwo
 \else \expandafter \@secondoftwo
 \fi
}%
\providecommand \natexlab [1]{#1}%
\providecommand \enquote  [1]{``#1''}%
\providecommand \bibnamefont  [1]{#1}%
\providecommand \bibfnamefont [1]{#1}%
\providecommand \citenamefont [1]{#1}%
\providecommand \href@noop [0]{\@secondoftwo}%
\providecommand \href [0]{\begingroup \@sanitize@url \@href}%
\providecommand \@href[1]{\@@startlink{#1}\@@href}%
\providecommand \@@href[1]{\endgroup#1\@@endlink}%
\providecommand \@sanitize@url [0]{\catcode `\\12\catcode `\$12\catcode
  `\&12\catcode `\#12\catcode `\^12\catcode `\_12\catcode `\%12\relax}%
\providecommand \@@startlink[1]{}%
\providecommand \@@endlink[0]{}%
\providecommand \url  [0]{\begingroup\@sanitize@url \@url }%
\providecommand \@url [1]{\endgroup\@href {#1}{\urlprefix }}%
\providecommand \urlprefix  [0]{URL }%
\providecommand \Eprint [0]{\href }%
\providecommand \doibase [0]{http://dx.doi.org/}%
\providecommand \selectlanguage [0]{\@gobble}%
\providecommand \bibinfo  [0]{\@secondoftwo}%
\providecommand \bibfield  [0]{\@secondoftwo}%
\providecommand \translation [1]{[#1]}%
\providecommand \BibitemOpen [0]{}%
\providecommand \bibitemStop [0]{}%
\providecommand \bibitemNoStop [0]{.\EOS\space}%
\providecommand \EOS [0]{\spacefactor3000\relax}%
\providecommand \BibitemShut  [1]{\csname bibitem#1\endcsname}%
\let\auto@bib@innerbib\@empty
\bibitem [{\citenamefont {Kwok}\ and\ \citenamefont
  {Untersteiner}(2011)}]{OneWatt}%
  \BibitemOpen
  \bibfield  {author} {\bibinfo {author} {\bibfnamefont {R.}~\bibnamefont
  {Kwok}}\ and\ \bibinfo {author} {\bibfnamefont {N.}~\bibnamefont
  {Untersteiner}},\ }\href@noop {} {\bibfield  {journal} {\bibinfo  {journal}
  {Phys. Today}\ }\textbf {\bibinfo {volume} {64}},\ \bibinfo {pages} {36}
  (\bibinfo {year} {2011})}\BibitemShut {NoStop}%
\bibitem [{\citenamefont {Kwok}\ \emph {et~al.}(2021)\citenamefont {Kwok},
  \citenamefont {Petty}, \citenamefont {Bagnardi}, \citenamefont {Kurtz},
  \citenamefont {Cunningham}, \citenamefont {Ivanoff},\ and\ \citenamefont
  {Kacimi}}]{Kwok:2021}%
  \BibitemOpen
  \bibfield  {author} {\bibinfo {author} {\bibfnamefont {R.}~\bibnamefont
  {Kwok}}, \bibinfo {author} {\bibfnamefont {A.~A.}\ \bibnamefont {Petty}},
  \bibinfo {author} {\bibfnamefont {M.}~\bibnamefont {Bagnardi}}, \bibinfo
  {author} {\bibfnamefont {N.~T.}\ \bibnamefont {Kurtz}}, \bibinfo {author}
  {\bibfnamefont {G.~F.}\ \bibnamefont {Cunningham}}, \bibinfo {author}
  {\bibfnamefont {A.}~\bibnamefont {Ivanoff}}, \ and\ \bibinfo {author}
  {\bibfnamefont {S.}~\bibnamefont {Kacimi}},\ }\href {\doibase
  10.5194/tc-15-821-2021} {\bibfield  {journal} {\bibinfo  {journal}
  {Cryosphere}\ }\textbf {\bibinfo {volume} {15}},\ \bibinfo {pages} {821}
  (\bibinfo {year} {2021})}\BibitemShut {NoStop}%
\bibitem [{\citenamefont {Rothrock}\ and\ \citenamefont
  {Thorndike}(1980)}]{Rothrock:1980}%
  \BibitemOpen
  \bibfield  {author} {\bibinfo {author} {\bibfnamefont {D.~A.}\ \bibnamefont
  {Rothrock}}\ and\ \bibinfo {author} {\bibfnamefont {A.~S.}\ \bibnamefont
  {Thorndike}},\ }\href@noop {} {\bibfield  {journal} {\bibinfo  {journal} {J.
  Geophys. Res.-Oceans}\ }\textbf {\bibinfo {volume} {85}},\ \bibinfo {pages}
  {3955} (\bibinfo {year} {1980})}\BibitemShut {NoStop}%
\bibitem [{\citenamefont {Rothrock}\ and\ \citenamefont
  {Thorndike}(1984)}]{Rothrock:1984}%
  \BibitemOpen
  \bibfield  {author} {\bibinfo {author} {\bibfnamefont {D.~A.}\ \bibnamefont
  {Rothrock}}\ and\ \bibinfo {author} {\bibfnamefont {A.~S.}\ \bibnamefont
  {Thorndike}},\ }\href@noop {} {\bibfield  {journal} {\bibinfo  {journal} {J.
  Geophys. Res.-Oceans}\ }\textbf {\bibinfo {volume} {89}},\ \bibinfo {pages}
  {6477} (\bibinfo {year} {1984})}\BibitemShut {NoStop}%
\bibitem [{\citenamefont {Manabe}\ and\ \citenamefont
  {Wetherald}(1975)}]{MW75}%
  \BibitemOpen
  \bibfield  {author} {\bibinfo {author} {\bibfnamefont {S.}~\bibnamefont
  {Manabe}}\ and\ \bibinfo {author} {\bibfnamefont {R.~T.}\ \bibnamefont
  {Wetherald}},\ }\href@noop {} {\bibfield  {journal} {\bibinfo  {journal} {J.
  Atmos. Sci.}\ }\textbf {\bibinfo {volume} {32}},\ \bibinfo {pages} {3}
  (\bibinfo {year} {1975})}\BibitemShut {NoStop}%
\bibitem [{\citenamefont {Untersteiner}\ \emph {et~al.}(2007)\citenamefont
  {Untersteiner}, \citenamefont {Thorndike}, \citenamefont {Rothrock},\ and\
  \citenamefont {Hunkins}}]{Untersteiner:2007}%
  \BibitemOpen
  \bibfield  {author} {\bibinfo {author} {\bibfnamefont {N.}~\bibnamefont
  {Untersteiner}}, \bibinfo {author} {\bibfnamefont {A.~S.}\ \bibnamefont
  {Thorndike}}, \bibinfo {author} {\bibfnamefont {D.~A.}\ \bibnamefont
  {Rothrock}}, \ and\ \bibinfo {author} {\bibfnamefont {K.~L.}\ \bibnamefont
  {Hunkins}},\ }\href@noop {} {\bibfield  {journal} {\bibinfo  {journal}
  {Arctic}\ }\textbf {\bibinfo {volume} {60}},\ \bibinfo {pages} {327}
  (\bibinfo {year} {2007})}\BibitemShut {NoStop}%
\bibitem [{\citenamefont {Rothrock}(1975)}]{Rothrock:1975}%
  \BibitemOpen
  \bibfield  {author} {\bibinfo {author} {\bibfnamefont {D.~A.}\ \bibnamefont
  {Rothrock}},\ }\href@noop {} {\bibfield  {journal} {\bibinfo  {journal}
  {Annu. Rev. Earth Planet Sci.}\ }\textbf {\bibinfo {volume} {3}},\ \bibinfo
  {pages} {317} (\bibinfo {year} {1975})}\BibitemShut {NoStop}%
\bibitem [{\citenamefont {Vella}\ and\ \citenamefont
  {Wettlaufer}(2008)}]{VW08}%
  \BibitemOpen
  \bibfield  {author} {\bibinfo {author} {\bibfnamefont {D.}~\bibnamefont
  {Vella}}\ and\ \bibinfo {author} {\bibfnamefont {J.~S.}\ \bibnamefont
  {Wettlaufer}},\ }\href@noop {} {\bibfield  {journal} {\bibinfo  {journal} {J.
  Geophys. Res.-Oceans}\ }\textbf {\bibinfo {volume} {113}},\ \bibinfo {pages}
  {C11011} (\bibinfo {year} {2008})}\BibitemShut {NoStop}%
\bibitem [{\citenamefont {Coon}\ \emph {et~al.}(2007)\citenamefont {Coon},
  \citenamefont {Kwok}, \citenamefont {Levy}, \citenamefont {Pruis},
  \citenamefont {Schreyer},\ and\ \citenamefont {Sulsky}}]{Coon:2007}%
  \BibitemOpen
  \bibfield  {author} {\bibinfo {author} {\bibfnamefont {M.}~\bibnamefont
  {Coon}}, \bibinfo {author} {\bibfnamefont {R.}~\bibnamefont {Kwok}}, \bibinfo
  {author} {\bibfnamefont {G.}~\bibnamefont {Levy}}, \bibinfo {author}
  {\bibfnamefont {M.}~\bibnamefont {Pruis}}, \bibinfo {author} {\bibfnamefont
  {H.}~\bibnamefont {Schreyer}}, \ and\ \bibinfo {author} {\bibfnamefont
  {D.}~\bibnamefont {Sulsky}},\ }\href {\doibase 10.1029/2005JC003393}
  {\bibfield  {journal} {\bibinfo  {journal} {J. Geophys. Res.-Oceans}\
  }\textbf {\bibinfo {volume} {112}},\ \bibinfo {pages} {C11S90} (\bibinfo
  {year} {2007})}\BibitemShut {NoStop}%
\bibitem [{\citenamefont {Feltham}(2008)}]{Feltham:2008}%
  \BibitemOpen
  \bibfield  {author} {\bibinfo {author} {\bibfnamefont {D.~L.}\ \bibnamefont
  {Feltham}},\ }\href {\doibase 10.1146/annurev.fluid.40.111406.102151}
  {\bibfield  {journal} {\bibinfo  {journal} {Annu. Rev. Fluid Mech.}\ }\textbf
  {\bibinfo {volume} {40}},\ \bibinfo {pages} {91} (\bibinfo {year}
  {2008})}\BibitemShut {NoStop}%
\bibitem [{\citenamefont {Roberts}\ \emph {et~al.}(2019)\citenamefont
  {Roberts}, \citenamefont {Hunke}, \citenamefont {Kamal}, \citenamefont
  {Lipscomb}, \citenamefont {Horvat},\ and\ \citenamefont
  {Maslowski}}]{Roberts:2019}%
  \BibitemOpen
  \bibfield  {author} {\bibinfo {author} {\bibfnamefont {A.~F.}\ \bibnamefont
  {Roberts}}, \bibinfo {author} {\bibfnamefont {E.~C.}\ \bibnamefont {Hunke}},
  \bibinfo {author} {\bibfnamefont {S.~M.}\ \bibnamefont {Kamal}}, \bibinfo
  {author} {\bibfnamefont {W.~H.}\ \bibnamefont {Lipscomb}}, \bibinfo {author}
  {\bibfnamefont {C.}~\bibnamefont {Horvat}}, \ and\ \bibinfo {author}
  {\bibfnamefont {W.}~\bibnamefont {Maslowski}},\ }\href {\doibase
  10.1029/2018MS001395} {\bibfield  {journal} {\bibinfo  {journal} {J. Adv.
  Model. Earth Syst.}\ }\textbf {\bibinfo {volume} {11}},\ \bibinfo {pages}
  {771} (\bibinfo {year} {2019})}\BibitemShut {NoStop}%
\bibitem [{\citenamefont {Thorndike}\ \emph {et~al.}(1975)\citenamefont
  {Thorndike}, \citenamefont {Rothrock}, \citenamefont {Maykut},\ and\
  \citenamefont {Colony}}]{Thorndike:1975}%
  \BibitemOpen
  \bibfield  {author} {\bibinfo {author} {\bibfnamefont {A.~S.}\ \bibnamefont
  {Thorndike}}, \bibinfo {author} {\bibfnamefont {D.~A.}\ \bibnamefont
  {Rothrock}}, \bibinfo {author} {\bibfnamefont {G.~A.}\ \bibnamefont
  {Maykut}}, \ and\ \bibinfo {author} {\bibfnamefont {R.}~\bibnamefont
  {Colony}},\ }\href@noop {} {\bibfield  {journal} {\bibinfo  {journal} {J.
  Geophys. Res.}\ }\textbf {\bibinfo {volume} {80}},\ \bibinfo {pages} {4501}
  (\bibinfo {year} {1975})}\BibitemShut {NoStop}%
\bibitem [{\citenamefont {Thorndike}(1992)}]{Thorndike:1992}%
  \BibitemOpen
  \bibfield  {author} {\bibinfo {author} {\bibfnamefont {A.~S.}\ \bibnamefont
  {Thorndike}},\ }\href@noop {} {\bibfield  {journal} {\bibinfo  {journal} {J.
  Geophys. Res.-Oceans}\ }\textbf {\bibinfo {volume} {97}},\ \bibinfo {pages}
  {12601} (\bibinfo {year} {1992})}\BibitemShut {NoStop}%
\bibitem [{\citenamefont {Thorndike}(2000)}]{Thorndike:2000}%
  \BibitemOpen
  \bibfield  {author} {\bibinfo {author} {\bibfnamefont {A.~S.}\ \bibnamefont
  {Thorndike}},\ }\href@noop {} {\bibfield  {journal} {\bibinfo  {journal} {J.
  Geophys. Res.-Oceans}\ }\textbf {\bibinfo {volume} {105}},\ \bibinfo {pages}
  {1311} (\bibinfo {year} {2000})}\BibitemShut {NoStop}%
\bibitem [{\citenamefont {Godlovitch}\ \emph {et~al.}(2011)\citenamefont
  {Godlovitch}, \citenamefont {Illner},\ and\ \citenamefont
  {Monahan}}]{godlovitch2011}%
  \BibitemOpen
  \bibfield  {author} {\bibinfo {author} {\bibfnamefont {D.}~\bibnamefont
  {Godlovitch}}, \bibinfo {author} {\bibfnamefont {R.}~\bibnamefont {Illner}},
  \ and\ \bibinfo {author} {\bibfnamefont {A.}~\bibnamefont {Monahan}},\
  }\href@noop {} {\bibfield  {journal} {\bibinfo  {journal} {J. Geophys.
  Res.-Oceans}\ }\textbf {\bibinfo {volume} {116}} (\bibinfo {year}
  {2011})}\BibitemShut {NoStop}%
\bibitem [{\citenamefont {Horvat}\ and\ \citenamefont
  {Tziperman}(2015)}]{Horvat:2015}%
  \BibitemOpen
  \bibfield  {author} {\bibinfo {author} {\bibfnamefont {C.}~\bibnamefont
  {Horvat}}\ and\ \bibinfo {author} {\bibfnamefont {E.}~\bibnamefont
  {Tziperman}},\ }\href {\doibase 10.5194/tc-9-2119-2015} {\bibfield  {journal}
  {\bibinfo  {journal} {Cryosphere}\ }\textbf {\bibinfo {volume} {9}},\
  \bibinfo {pages} {2119} (\bibinfo {year} {2015})}\BibitemShut {NoStop}%
\bibitem [{\citenamefont {Toppaladoddi}\ and\ \citenamefont
  {Wettlaufer}(2017)}]{TW2017}%
  \BibitemOpen
  \bibfield  {author} {\bibinfo {author} {\bibfnamefont {S.}~\bibnamefont
  {Toppaladoddi}}\ and\ \bibinfo {author} {\bibfnamefont {J.~S.}\ \bibnamefont
  {Wettlaufer}},\ }\href@noop {} {\bibfield  {journal} {\bibinfo  {journal} {J.
  Stat. Phys.}\ }\textbf {\bibinfo {volume} {167}},\ \bibinfo {pages} {683}
  (\bibinfo {year} {2017})}\BibitemShut {NoStop}%
\bibitem [{\citenamefont {Toppaladoddi}\ and\ \citenamefont
  {Wettlaufer}(2015)}]{TW2015}%
  \BibitemOpen
  \bibfield  {author} {\bibinfo {author} {\bibfnamefont {S.}~\bibnamefont
  {Toppaladoddi}}\ and\ \bibinfo {author} {\bibfnamefont {J.~S.}\ \bibnamefont
  {Wettlaufer}},\ }\href@noop {} {\bibfield  {journal} {\bibinfo  {journal}
  {Phys. Rev. Lett.}\ }\textbf {\bibinfo {volume} {115}},\ \bibinfo {pages}
  {148501} (\bibinfo {year} {2015})}\BibitemShut {NoStop}%
\bibitem [{\citenamefont {Pawula}(1967)}]{Pawula}%
  \BibitemOpen
  \bibfield  {author} {\bibinfo {author} {\bibfnamefont {R.~F.}\ \bibnamefont
  {Pawula}},\ }\href@noop {} {\bibfield  {journal} {\bibinfo  {journal} {Phys.
  Rev.}\ }\textbf {\bibinfo {volume} {162}},\ \bibinfo {pages} {186} (\bibinfo
  {year} {1967})}\BibitemShut {NoStop}%
\bibitem [{\citenamefont {Courant}\ and\ \citenamefont
  {Hilbert}(1953)}]{Courant}%
  \BibitemOpen
  \bibfield  {author} {\bibinfo {author} {\bibfnamefont {R.}~\bibnamefont
  {Courant}}\ and\ \bibinfo {author} {\bibfnamefont {D.}~\bibnamefont
  {Hilbert}},\ }\href@noop {} {\emph {\bibinfo {title} {Methods of Mathematical
  Physics}}},\ Vol.~\bibinfo {volume} {II}\ (\bibinfo  {publisher}
  {Interscience},\ \bibinfo {address} {New York, NY},\ \bibinfo {year}
  {1953})\BibitemShut {NoStop}%
\bibitem [{\citenamefont {Agarwal}\ and\ \citenamefont
  {Wettlaufer}(2017)}]{agarwal2017}%
  \BibitemOpen
  \bibfield  {author} {\bibinfo {author} {\bibfnamefont {S.}~\bibnamefont
  {Agarwal}}\ and\ \bibinfo {author} {\bibfnamefont {J.~S.}\ \bibnamefont
  {Wettlaufer}},\ }\href@noop {} {\bibfield  {journal} {\bibinfo  {journal} {J.
  Clim.}\ }\textbf {\bibinfo {volume} {30}},\ \bibinfo {pages} {4873} (\bibinfo
  {year} {2017})}\BibitemShut {NoStop}%
\bibitem [{\citenamefont {Kwok}\ \emph {et~al.}(2009)\citenamefont {Kwok},
  \citenamefont {Cunningham}, \citenamefont {Wensnahan}, \citenamefont {Rigor},
  \citenamefont {Zwally},\ and\ \citenamefont {Yi}}]{kwok2009}%
  \BibitemOpen
  \bibfield  {author} {\bibinfo {author} {\bibfnamefont {R.}~\bibnamefont
  {Kwok}}, \bibinfo {author} {\bibfnamefont {G.}~\bibnamefont {Cunningham}},
  \bibinfo {author} {\bibfnamefont {M.}~\bibnamefont {Wensnahan}}, \bibinfo
  {author} {\bibfnamefont {I.}~\bibnamefont {Rigor}}, \bibinfo {author}
  {\bibfnamefont {H.}~\bibnamefont {Zwally}}, \ and\ \bibinfo {author}
  {\bibfnamefont {D.}~\bibnamefont {Yi}},\ }\href@noop {} {\bibfield  {journal}
  {\bibinfo  {journal} {J. Geophys. Res.-Oceans}\ }\textbf {\bibinfo {volume}
  {114}} (\bibinfo {year} {2009})}\BibitemShut {NoStop}%
\bibitem [{\citenamefont {Kwok}\ and\ \citenamefont
  {Cunningham}(2015)}]{kwok2015}%
  \BibitemOpen
  \bibfield  {author} {\bibinfo {author} {\bibfnamefont {R.}~\bibnamefont
  {Kwok}}\ and\ \bibinfo {author} {\bibfnamefont {G.}~\bibnamefont
  {Cunningham}},\ }\href@noop {} {\bibfield  {journal} {\bibinfo  {journal}
  {Phil. Trans. R. Soc. A}\ }\textbf {\bibinfo {volume} {373}},\ \bibinfo
  {pages} {20140157} (\bibinfo {year} {2015})}\BibitemShut {NoStop}%
\bibitem [{\citenamefont {Semtner}(1976)}]{semtner1976}%
  \BibitemOpen
  \bibfield  {author} {\bibinfo {author} {\bibfnamefont {A.~J.}\ \bibnamefont
  {Semtner}},\ }\href@noop {} {\bibfield  {journal} {\bibinfo  {journal} {J.
  Phys. Oceanogr.}\ }\textbf {\bibinfo {volume} {6}},\ \bibinfo {pages} {379}
  (\bibinfo {year} {1976})}\BibitemShut {NoStop}%
\bibitem [{\citenamefont {Maykut}\ and\ \citenamefont
  {Untersteiner}(1971)}]{MU71}%
  \BibitemOpen
  \bibfield  {author} {\bibinfo {author} {\bibfnamefont {G.~A.}\ \bibnamefont
  {Maykut}}\ and\ \bibinfo {author} {\bibfnamefont {N.}~\bibnamefont
  {Untersteiner}},\ }\href@noop {} {\bibfield  {journal} {\bibinfo  {journal}
  {J. Geophys. Res.}\ }\textbf {\bibinfo {volume} {76}},\ \bibinfo {pages}
  {1550 } (\bibinfo {year} {1971})}\BibitemShut {NoStop}%
\bibitem [{\citenamefont {Eisenman}\ and\ \citenamefont
  {Wettlaufer}(2009)}]{EW09}%
  \BibitemOpen
  \bibfield  {author} {\bibinfo {author} {\bibfnamefont {I.}~\bibnamefont
  {Eisenman}}\ and\ \bibinfo {author} {\bibfnamefont {J.~S.}\ \bibnamefont
  {Wettlaufer}},\ }\href@noop {} {\bibfield  {journal} {\bibinfo  {journal}
  {Proc. Natl. Acad. Sci. USA}\ }\textbf {\bibinfo {volume} {106}},\ \bibinfo
  {pages} {28} (\bibinfo {year} {2009})}\BibitemShut {NoStop}%
\bibitem [{\citenamefont {Bitz}\ and\ \citenamefont
  {Lipscomb}(1999)}]{bitz1999}%
  \BibitemOpen
  \bibfield  {author} {\bibinfo {author} {\bibfnamefont {C.~M.}\ \bibnamefont
  {Bitz}}\ and\ \bibinfo {author} {\bibfnamefont {W.~H.}\ \bibnamefont
  {Lipscomb}},\ }\href@noop {} {\bibfield  {journal} {\bibinfo  {journal} {J.
  Geophys. Res.-Oceans}\ }\textbf {\bibinfo {volume} {104}},\ \bibinfo {pages}
  {15669} (\bibinfo {year} {1999})}\BibitemShut {NoStop}%
\bibitem [{\citenamefont {Yu}\ \emph {et~al.}(2004)\citenamefont {Yu},
  \citenamefont {Maykut},\ and\ \citenamefont {Rothrock}}]{yu2004}%
  \BibitemOpen
  \bibfield  {author} {\bibinfo {author} {\bibfnamefont {Y.}~\bibnamefont
  {Yu}}, \bibinfo {author} {\bibfnamefont {G.}~\bibnamefont {Maykut}}, \ and\
  \bibinfo {author} {\bibfnamefont {D.}~\bibnamefont {Rothrock}},\ }\href@noop
  {} {\bibfield  {journal} {\bibinfo  {journal} {J. Geophys. Res.-Oceans}\
  }\textbf {\bibinfo {volume} {109}} (\bibinfo {year} {2004})}\BibitemShut
  {NoStop}%
\bibitem [{\citenamefont {Landy}\ \emph {et~al.}(2022)\citenamefont {Landy},
  \citenamefont {Dawson}, \citenamefont {Tsamados}, \citenamefont {Bushuk},
  \citenamefont {Stroeve}, \citenamefont {Howell}, \citenamefont {Krumpen},
  \citenamefont {Babb}, \citenamefont {Komarov}, \citenamefont {Heorton} \emph
  {et~al.}}]{Landy2022}%
  \BibitemOpen
  \bibfield  {author} {\bibinfo {author} {\bibfnamefont {J.~C.}\ \bibnamefont
  {Landy}}, \bibinfo {author} {\bibfnamefont {G.~J.}\ \bibnamefont {Dawson}},
  \bibinfo {author} {\bibfnamefont {M.}~\bibnamefont {Tsamados}}, \bibinfo
  {author} {\bibfnamefont {M.}~\bibnamefont {Bushuk}}, \bibinfo {author}
  {\bibfnamefont {J.~C.}\ \bibnamefont {Stroeve}}, \bibinfo {author}
  {\bibfnamefont {S.~E.~L.}\ \bibnamefont {Howell}}, \bibinfo {author}
  {\bibfnamefont {T.}~\bibnamefont {Krumpen}}, \bibinfo {author} {\bibfnamefont
  {D.~G.}\ \bibnamefont {Babb}}, \bibinfo {author} {\bibfnamefont {A.~S.}\
  \bibnamefont {Komarov}}, \bibinfo {author} {\bibfnamefont {H.~D. B.~S.}\
  \bibnamefont {Heorton}},  \emph {et~al.},\ }\href@noop {} {\bibfield
  {journal} {\bibinfo  {journal} {Nature}\ }\textbf {\bibinfo {volume} {609}},\
  \bibinfo {pages} {517} (\bibinfo {year} {2022})}\BibitemShut {NoStop}%
\bibitem [{\citenamefont {Wettlaufer}(1991)}]{JSW91}%
  \BibitemOpen
  \bibfield  {author} {\bibinfo {author} {\bibfnamefont {J.~S.}\ \bibnamefont
  {Wettlaufer}},\ }\href@noop {} {\bibfield  {journal} {\bibinfo  {journal} {J.
  Geophys. Res.-Oceans}\ }\textbf {\bibinfo {volume} {96}},\ \bibinfo {pages}
  {7215} (\bibinfo {year} {1991})}\BibitemShut {NoStop}%
\bibitem [{\citenamefont {McPhee}(1992)}]{mcphee1992}%
  \BibitemOpen
  \bibfield  {author} {\bibinfo {author} {\bibfnamefont {M.~G.}\ \bibnamefont
  {McPhee}},\ }\href@noop {} {\bibfield  {journal} {\bibinfo  {journal} {J.
  Geophys. Res.-Oceans}\ }\textbf {\bibinfo {volume} {97}},\ \bibinfo {pages}
  {5365} (\bibinfo {year} {1992})}\BibitemShut {NoStop}%
\bibitem [{\citenamefont {Stanton}\ \emph {et~al.}(2012)\citenamefont
  {Stanton}, \citenamefont {Shaw},\ and\ \citenamefont
  {Hutchings}}]{stanton2012}%
  \BibitemOpen
  \bibfield  {author} {\bibinfo {author} {\bibfnamefont {T.~P.}\ \bibnamefont
  {Stanton}}, \bibinfo {author} {\bibfnamefont {W.~J.}\ \bibnamefont {Shaw}}, \
  and\ \bibinfo {author} {\bibfnamefont {J.~K.}\ \bibnamefont {Hutchings}},\
  }\href@noop {} {\bibfield  {journal} {\bibinfo  {journal} {J. Geophys.
  Res.-Oceans}\ }\textbf {\bibinfo {volume} {117}} (\bibinfo {year}
  {2012})}\BibitemShut {NoStop}%
\bibitem [{\citenamefont {Timmermans}(2015)}]{MLT2015}%
  \BibitemOpen
  \bibfield  {author} {\bibinfo {author} {\bibfnamefont {M.-L.}\ \bibnamefont
  {Timmermans}},\ }\href@noop {} {\bibfield  {journal} {\bibinfo  {journal}
  {Geophys. Res. Lett.}\ }\textbf {\bibinfo {volume} {42}},\ \bibinfo {pages}
  {6399} (\bibinfo {year} {2015})}\BibitemShut {NoStop}%
\bibitem [{\citenamefont {Timmermans}\ \emph {et~al.}(2018)\citenamefont
  {Timmermans}, \citenamefont {Toole},\ and\ \citenamefont
  {Krishfield}}]{MLT2018}%
  \BibitemOpen
  \bibfield  {author} {\bibinfo {author} {\bibfnamefont {M.-L.}\ \bibnamefont
  {Timmermans}}, \bibinfo {author} {\bibfnamefont {J.}~\bibnamefont {Toole}}, \
  and\ \bibinfo {author} {\bibfnamefont {R.}~\bibnamefont {Krishfield}},\
  }\href@noop {} {\bibfield  {journal} {\bibinfo  {journal} {Sci. Adv.}\
  }\textbf {\bibinfo {volume} {4}},\ \bibinfo {pages} {eaat6773} (\bibinfo
  {year} {2018})}\BibitemShut {NoStop}%
\bibitem [{\citenamefont {Hibler~III}(1979)}]{hibler1979}%
  \BibitemOpen
  \bibfield  {author} {\bibinfo {author} {\bibfnamefont {W.~D.}\ \bibnamefont
  {Hibler~III}},\ }\href@noop {} {\bibfield  {journal} {\bibinfo  {journal} {J.
  Phys. Oceanogr.}\ }\textbf {\bibinfo {volume} {9}},\ \bibinfo {pages} {815}
  (\bibinfo {year} {1979})}\BibitemShut {NoStop}%
\bibitem [{\citenamefont {Bitz}\ \emph {et~al.}(2001)\citenamefont {Bitz},
  \citenamefont {Holland}, \citenamefont {Weaver},\ and\ \citenamefont
  {Eby}}]{bitz2001}%
  \BibitemOpen
  \bibfield  {author} {\bibinfo {author} {\bibfnamefont {C.~M.}\ \bibnamefont
  {Bitz}}, \bibinfo {author} {\bibfnamefont {M.~M.}\ \bibnamefont {Holland}},
  \bibinfo {author} {\bibfnamefont {A.~J.}\ \bibnamefont {Weaver}}, \ and\
  \bibinfo {author} {\bibfnamefont {M.}~\bibnamefont {Eby}},\ }\href@noop {}
  {\bibfield  {journal} {\bibinfo  {journal} {J. Geophys. Res.-Oceans}\
  }\textbf {\bibinfo {volume} {106}},\ \bibinfo {pages} {2441} (\bibinfo {year}
  {2001})}\BibitemShut {NoStop}%
\bibitem [{\citenamefont {Smith}\ \emph {et~al.}(2022)\citenamefont {Smith},
  \citenamefont {Holland}, \citenamefont {Petty}, \citenamefont {Light},\ and\
  \citenamefont {Bailey}}]{smith2022}%
  \BibitemOpen
  \bibfield  {author} {\bibinfo {author} {\bibfnamefont {M.~M.}\ \bibnamefont
  {Smith}}, \bibinfo {author} {\bibfnamefont {M.~M.}\ \bibnamefont {Holland}},
  \bibinfo {author} {\bibfnamefont {A.~A.}\ \bibnamefont {Petty}}, \bibinfo
  {author} {\bibfnamefont {B.}~\bibnamefont {Light}}, \ and\ \bibinfo {author}
  {\bibfnamefont {D.~A.}\ \bibnamefont {Bailey}},\ }\href@noop {} {\bibfield
  {journal} {\bibinfo  {journal} {J. Geophys. Res.-Oceans}\ ,\ \bibinfo {pages}
  {e2022JC019044}} (\bibinfo {year} {2022})}\BibitemShut {NoStop}%
\bibitem [{\citenamefont {Castro-Morales}\ \emph {et~al.}(2014)\citenamefont
  {Castro-Morales}, \citenamefont {Kauker}, \citenamefont {Losch},
  \citenamefont {Hendricks}, \citenamefont {Riemann-Campe},\ and\ \citenamefont
  {Gerdes}}]{castro2014}%
  \BibitemOpen
  \bibfield  {author} {\bibinfo {author} {\bibfnamefont {K.}~\bibnamefont
  {Castro-Morales}}, \bibinfo {author} {\bibfnamefont {F.}~\bibnamefont
  {Kauker}}, \bibinfo {author} {\bibfnamefont {M.}~\bibnamefont {Losch}},
  \bibinfo {author} {\bibfnamefont {S.}~\bibnamefont {Hendricks}}, \bibinfo
  {author} {\bibfnamefont {K.}~\bibnamefont {Riemann-Campe}}, \ and\ \bibinfo
  {author} {\bibfnamefont {R.}~\bibnamefont {Gerdes}},\ }\href@noop {}
  {\bibfield  {journal} {\bibinfo  {journal} {J. Geophys. Res.-Oceans}\
  }\textbf {\bibinfo {volume} {119}},\ \bibinfo {pages} {559} (\bibinfo {year}
  {2014})}\BibitemShut {NoStop}%
\bibitem [{\citenamefont {Stroeve}\ \emph {et~al.}(2014)\citenamefont
  {Stroeve}, \citenamefont {Barrett}, \citenamefont {Serreze},\ and\
  \citenamefont {Schweiger}}]{stroeve2014}%
  \BibitemOpen
  \bibfield  {author} {\bibinfo {author} {\bibfnamefont {J.}~\bibnamefont
  {Stroeve}}, \bibinfo {author} {\bibfnamefont {A.}~\bibnamefont {Barrett}},
  \bibinfo {author} {\bibfnamefont {M.}~\bibnamefont {Serreze}}, \ and\
  \bibinfo {author} {\bibfnamefont {A.}~\bibnamefont {Schweiger}},\ }\href@noop
  {} {\bibfield  {journal} {\bibinfo  {journal} {Cryosphere}\ }\textbf
  {\bibinfo {volume} {8}},\ \bibinfo {pages} {1839} (\bibinfo {year}
  {2014})}\BibitemShut {NoStop}%
\bibitem [{\citenamefont {Agarwal}\ and\ \citenamefont
  {Wettlaufer}(2018)}]{agarwal2018}%
  \BibitemOpen
  \bibfield  {author} {\bibinfo {author} {\bibfnamefont {S.}~\bibnamefont
  {Agarwal}}\ and\ \bibinfo {author} {\bibfnamefont {J.~S.}\ \bibnamefont
  {Wettlaufer}},\ }\href@noop {} {\bibfield  {journal} {\bibinfo  {journal}
  {Phil. Trans. R. Soc. A}\ }\textbf {\bibinfo {volume} {376}},\ \bibinfo
  {pages} {20170332} (\bibinfo {year} {2018})}\BibitemShut {NoStop}%
\bibitem [{\citenamefont {Wei}\ \emph {et~al.}(2020)\citenamefont {Wei},
  \citenamefont {Yan}, \citenamefont {Qi}, \citenamefont {Ding},\ and\
  \citenamefont {Wang}}]{wei2020}%
  \BibitemOpen
  \bibfield  {author} {\bibinfo {author} {\bibfnamefont {T.}~\bibnamefont
  {Wei}}, \bibinfo {author} {\bibfnamefont {Q.}~\bibnamefont {Yan}}, \bibinfo
  {author} {\bibfnamefont {W.}~\bibnamefont {Qi}}, \bibinfo {author}
  {\bibfnamefont {M.}~\bibnamefont {Ding}}, \ and\ \bibinfo {author}
  {\bibfnamefont {C.}~\bibnamefont {Wang}},\ }\href@noop {} {\bibfield
  {journal} {\bibinfo  {journal} {Environ. Res. Lett.}\ }\textbf {\bibinfo
  {volume} {15}},\ \bibinfo {pages} {104079} (\bibinfo {year}
  {2020})}\BibitemShut {NoStop}%
\end{thebibliography}%
\end{document}